\documentclass[linenumbers,preprint,tightenlines,preprintnumbers,aps,eqsecnum,nofootinbib,superscriptaddress,showpacs,floatfix]{revtex4}

\usepackage{amsmath,amssymb,amsbsy,amsfonts,bm}
\usepackage{graphicx}
\usepackage{bm}
\usepackage{capt-of}
\usepackage[mathscr]{euscript}
\usepackage{color}
\usepackage{rotating}
\usepackage{slashed}
\usepackage{float}
\usepackage{hyperref}

\usepackage{setspace}
\graphicspath{{./figures/}}

\newcommand{\order}{{\it O}}
\newcommand{\ba}{\begin{eqnarray}}
\newcommand{\ea}{\end{eqnarray}}
\newcommand{\be} {\begin{equation}}
\newcommand{\ee} {\end{equation}}

\newcommand{\cpt}{\raise0.4ex\hbox{$\chi$}PT}
\newcommand{\scpt}{S\raise0.4ex\hbox{$\chi$}PT}
\newcommand{\rscpt}{rS\raise0.4ex\hbox{$\chi$}PT}

\newcommand{\cM}{\ensuremath{\mathcal{M}}}
\newcommand{\cMI}{\ensuremath{\mathcal{M}_I}}
\newcommand{\cMV}{\ensuremath{\mathcal{M}_V}}

\begin{document}

\singlespacing

\preprint{FERMILAB-PUB-12-651-PPD}

\title{Kaon semileptonic vector form factor and determination of 
$\vert V_{us}\vert$ using staggered fermions}

\author{A.~Bazavov}
\affiliation{Physics Department, Brookhaven National Laboratory, Upton, NY, USA}

\author{C.~Bernard}
\affiliation{Department of Physics, Washington University, St.~Louis, Missouri, USA}

\author{C.M.~Bouchard}
\affiliation{Department of Physics, The Ohio State University, Columbus, Ohio, USA}

\author{C.~DeTar}
\affiliation{Physics Department, University of Utah, Salt Lake City, Utah, USA}

\author{Daping~Du}
\affiliation{Physics Department, University of Illinois, Urbana, Illinois, USA}

\author{A.X.~El-Khadra}
\affiliation{Physics Department, University of Illinois, Urbana, Illinois, USA}

\author{J.~Foley}
\affiliation{Physics Department, University of Utah, Salt Lake City, Utah, USA}

\author{E.D.~Freeland}
\affiliation{Department of Physics, Benedictine University, Lisle, Illinois, USA}

\author{E.~G\'amiz}
\email{megamiz@ugr.es}
\affiliation{CAFPE and Departamento de F\'{\i}sica Te\'orica y del Cosmos,
Universidad de Granada, Granada, Spain}

\author{Steven~Gottlieb}
\affiliation{Department of Physics, Indiana University, Bloomington, Indiana, USA}

\author{U.M.~Heller}
\affiliation{American Physical Society, Ridge, New York, USA}

\author{Jongjeong Kim}
\affiliation{Department of Physics, University of Arizona, Tucson, Arizona, USA}

\author{A.S.~Kronfeld}
\affiliation{Fermi National Accelerator Laboratory, Batavia, Illinois, USA}

\author{J.~Laiho}
\affiliation{SUPA, School of Physics and Astronomy, University of Glasgow, Glasgow, UK}

\author{L.~Levkova}
\affiliation{Physics Department, University of Utah, Salt Lake City, Utah, USA}

\author{P.B.~Mackenzie}
\affiliation{Fermi National Accelerator Laboratory, Batavia, Illinois, USA}

\author{E.T.~Neil}
\affiliation{Fermi National Accelerator Laboratory, Batavia, Illinois, USA}

\author{M.B.~Oktay}
\affiliation{Physics Department, University of Utah, Salt Lake City, Utah, USA}

\author{Si-Wei Qiu}
\affiliation{Physics Department, University of Utah, Salt Lake City, Utah, USA}

\author{J.N.~Simone}
\affiliation{Fermi National Accelerator Laboratory, Batavia, Illinois, USA}

\author{R.~Sugar}
\affiliation{Department of Physics, University of California, Santa Barbara, California, USA}

\author{D.~Toussaint}
\affiliation{Department of Physics, University of Arizona, Tucson, Arizona, USA}

\author{R.S.~Van~de~Water}
\affiliation{Fermi National Accelerator Laboratory, Batavia, Illinois, USA}

\author{Ran Zhou}
\affiliation{Department of Physics, Indiana University, Bloomington, Indiana, USA}

\collaboration{Fermilab Lattice and MILC Collaborations}
\noaffiliation

\date{\today}

\begin{abstract}

Using staggered fermions and partially twisted boundary conditions, 
we calculate the 
$K$ meson semileptonic decay vector form factor at zero momentum transfer. The HISQ 
formulation is used for the valence quarks, while the sea quarks 
are simulated with the asqtad action (MILC $N_f=2+1$ configurations). For the 
chiral and continuum extrapolation we use two-loop continuum $\chi$PT,
supplemented by partially quenched staggered $\chi$PT at one loop. 
Our result is $f_+^{K\pi}(0) = 0.9667\pm 0.0023\pm0.0033$, where the first 
error is statistical and the second is the sum in quadrature of the systematic 
uncertainties. This result is the first $N_f=2+1$ calculation with two 
lattice spacings and a controlled continuum extrapolation. It is also 
the most precise result to date for the vector form factor 
and, although the central value is larger than previous unquenched lattice calculations, 
it is compatible with them within errors. Combining our value for $f_+^{K\pi}(0)$ 
with the latest experimental measurements of $K$ semileptonic decays, we obtain 
$\vert V_{us}\vert = 0.2238\pm0.0009\pm0.0005$, where the first error is from 
$f_+^{K\pi}(0)$ and the second one is experimental. As a byproduct of our calculation, we  
obtain the combination of low-energy constants 
$(C_{12}^r+C_{34}^r-(L_5^r)^2)(M_\rho) = (3.62\pm1.00)\times10^{-6}$.  

\end{abstract}

\pacs{12.15.Hh,12.38.Ge,13.20.Eb}

\maketitle

\section{Introduction}

\label{introduccion}

The Standard Model (SM) predicts that the matrix that describes the mixing 
between the different quark flavors, the Cabibbo-Kobayashi-Maskawa (CKM) 
matrix, has to be unitary. Any deviation from unitarity would indicate the 
existence of beyond-the-Standard-Model physics, so checking CKM 
unitarity is very important in the search for new physics (NP). In 
particular, tests of the unitarity of the first row of the CKM matrix can 
establish stringent constraints on the scale of allowed NP, even 
if unitarity is fulfilled. Given current experimental and 
theoretical inputs, the low-energy constraint coming from these tests 
($\Lambda > 11~{\rm TeV}$, where $\Lambda$ is the scale of NP)~\cite{Cirigliano10} 
is at the same level as those from $Z$-pole measurements~\cite{Cirigliano11}.

The two most precise grounds for the determination of the Cabibbo-Kobayashi-Maskawa 
matrix element $\vert V_{us}\vert$ are leptonic and semileptonic decays, with
competing errors at the 0.6\% level~\cite{Cirigliano11}. Determinations of 
$\vert V_{us}\vert$ from
hadronic $\tau$ decays have the potential of being competitive with the above  
determinations, but they are currently limited by uncertainties in the 
experimental data~\cite{tauHFAG}.

Both leptonic and semileptonic determinations require nonperturbative inputs, which 
can be precisely calculated on the lattice. In the case of leptonic decays, the ratio 
of the decay rates $K\to\mu\nu$ and $\pi\to\mu\nu$ can be combined with the ratio of decay 
constants $f_K/f_\pi$, calculated on the lattice with a precision better than 
$0.5\%$ for the averaged value~\cite{fKoverfpi1,fKoverfpi2,Aubin:2004fs,fKoverfpi3,fKoverfpi4,fKoverfpi5,fKoverfpi6,Bazavov:2013cp,FLAG,lataver}, 
to extract $\left\vert V_{us}/V_{ud}\right\vert$~\cite{Marciano04}. 
For semileptonic decays, the photon-inclusive decay rate for all $K\to\pi l\nu$ decay 
modes can be related to the CKM matrix element $\vert V_{us}\vert$ via Eq.~(4.37) of 
Ref.~\cite{Cirigliano11}
\ba\label{eq:Kl3def}
\Gamma_{K_{l3 (\gamma)}} = \frac{G_F^2M_K^5 C_K^2}{128\pi^3}S_{{\rm EW}}
\vert V_{us}f_+^{K^0\pi^-}(0)\vert^2 I_{Kl}^{(0)} 
\left(1+\delta_{{\rm EM}}^{Kl} + \delta_{{\rm SU(2)}}^{K\pi}\right)\,,
\ea
where the Clebsch-Gordon coefficient $C_K$ is equal to $1$ or $1/\sqrt{2}$ 
for neutral and charged kaons, respectively; $S_{{\rm EW}}=1.0223(5)$ is the short-distance 
universal electroweak correction; and, $I_{Kl}^{(0)}$ is a phase space 
integral which depends on the shape of the form factors $f_{\pm}^{K\pi}$. 
The parameters $\delta_{{\rm EM}}^{Kl}$ and $\delta_{{\rm SU(2)}}^{K\pi}$ contain long-distance 
electromagnetic and strong isospin-breaking corrections, respectively, and are defined 
as corrections to the $K^0$ mode. 
A discussion of the values and different determinations of those parameters can 
be found in Ref.~\cite{Cirigliano11}. We note that a different normalization from that of 
Ref.~\cite{Cirigliano11} is often used 
in the literature for the phase space integral~$I_{Kl}^{(0)}$, in which case the factor 
128 in Eq.~(\ref{eq:Kl3def}) is replaced by a factor of 192. 
 
The vector form factor $f_{+}^{K\pi}$, together with the scalar form factor 
$f_0^{K\pi}$, describes the hadronic matrix element of a vector current
\ba\label{eq:formfac}
\langle \pi (p_\pi)\vert V^\mu \vert K(p_K)\rangle =
f_+^{K\pi}(q^2) \left[p_K^\mu + p_\pi^\mu - \frac{m_K^2-m_\pi^2}{q^2}
q^\mu\right]+f_0^{K\pi}(q^2)\frac{m_K^2-m_\pi^2}{q^2}q^\mu\,,
\ea
where $q=p_K-p_\pi$ is the momentum transfer and $V^\mu=\bar s \gamma^\mu u$ 
is the appropriate flavor changing vector current. 
It is convenient to factor out $f_+^{K^0\pi^-}$ 
in (\ref{eq:Kl3def}) and normalize all experimental measurements to this channel. 
From now on, we will denote $f_+^{K^0\pi^-}(q^2)$ by $f_+(q^2)$ and 
$f_0^{K^0\pi^-}(q^2)$ by $f_0(q^2)$. Thus, by definition, 
$f_+(0)$ is given by the hadronic matrix element (\ref{eq:formfac}) for 
the $K^0\pi^-$ channel with meson masses set equal to the physical ones. 

The error associated with the lattice-QCD determination of $f_+(q^2=0)$ 
is around 0.5\%, but it is still the dominant uncertainty in the
extraction of $\vert V_{us}\vert$ from experimental data on $K$ semileptonic
decays. References~\cite{Antonellietal2010,MoulsonCKM12} quote 
$\vert V_{us}\vert f_+(0) = 0.2163(5)$, an experimental error of 0.23\%.  
Improvement in the determination of the form factor is thus crucial
in order to extract all the information from the available experimental data. 

In the search for deviations from SM predictions, one can also compare the values of 
$\vert V_{us}\vert$ as extracted from helicity-allowed semileptonic decays and 
helicity-suppressed leptonic decays. In particular, it is useful to study the ratio
\ba \label{eq:Rmu23}
R_{\mu 23} = \left(\frac{f_+^{K\pi}(0)}{f_K/f_\pi}\right)
\left(\left|\frac{V_{us}}{V_{ud}}\right|\frac{f_K}{f_\pi}\right)_{\mu 2}\,
\vert V_{ud}\vert\,\frac{1}{\left\lbrack\vert V_{us}\vert
f_+^{K\pi}(0)\right\rbrack_{l3}}\, ,
\ea
where the subscripts $\mu 2$ and $l 3$ indicate that these quantities are
obtained from experimental data on leptonic $K_{\mu 2}$ and semileptonic $K_{l 3}$
decays, respectively, and the corresponding SM formulae. 
The ratio in Eq.~(\ref{eq:Rmu23}) is unity in the SM by construction, but not 
in some extensions of the SM, such as those
with a charged Higgs. Again, the error in the current value $R_{\mu 23} = 0.999(7)$
\cite{Antonellietal2010} using the average of $\vert V_{ud}\vert$ from 
superallowed nuclear beta decays in Ref.~\cite{Vud08}, 
is limited by the precision of lattice-QCD inputs.

In this paper, we report on a lattice-QCD calculation of the form factor
$f_+(0)$ using the highly-improved staggered quark (HISQ)
action~\cite{Hisqaction} for valence quarks and the asqtad-improved staggered 
action~\cite{lepage-1999-59} for sea quarks. The goal of this analysis is to 
provide, using the staggered formulation, a determination of this parameter that is 
competitive with other state-of-the-art unquenched determinations~\cite{Boyle:2010bh,f+ETMC}. 
In order to avoid the complications associated with the fact that the simplest  
staggered vector currents are not conserved and need to be renormalized, 
we use the method developed by the HPQCD Collaboration in Ref.~\cite{HPQCD_Dtopi} and described 
in Sec.~\ref{sec:methodology}. This method uses correlation functions of a 
scalar current, which do not need a renormalization factor. 
Some tests of the general methodology and specific aspects such as the use of
random wall sources, the stability of the correlator fits, or the chiral and
continuum extrapolations  were already discussed in Refs.~\cite{latt2010,latt2011,latt2012}. 
Preliminary results for $f_+(0)$
using different light-quark actions, such as domain wall (DW) or overlap,
can be found in Refs.~\cite{Kanekolat12} and~\cite{Sivalingamlat12}, respectively. 
An even more precise calculation using HISQ valence quarks on HISQ configurations is 
already underway~\cite{latt2012}. The result presented here, however, is the 
first $N_f=2+1$ determination of $f_+(0)$ from multiple lattice spacings with a 
controlled continuum extrapolation, and the errors are already smaller than 
those of other recently published calculations~\cite{Boyle:2010bh,f+ETMC}.

This paper is organized as follows. In Sec.~\ref{sec:methodology}, we describe the 
main characteristics of our approach, and in Sec.~\ref{sec:numerical} the simulation 
details as well as the fitting strategy for the correlator fits.  
Section~\ref{sec:chiralfits} is dedicated to the chiral and continuum extrapolation, 
and Sec.\ref{sec:Errors} compiles the discussions on the different sources of 
systematic errors in our calculation. In Sec.~\ref{sec:p6LECs}, we give the 
result for the $\order(p^6)$ low-energy constants relevant for the 
calculation of $f_+(0)$. Finally, 
we present our final results and conclusions in Sec.~\ref{sec:Results}.

\section{Methodology}

\label{sec:methodology}

One of the main components of our analysis, which reduces both systematic and 
statistical errors, is to use the method developed by the HPQCD 
collaboration to study charm semileptonic decays \cite{HPQCD_Dtopi}. 
This method is based on the Ward identity relating the matrix 
element of a vector current to that of the corresponding scalar current  
\ba\label{eq:WI}
q^\mu\langle \pi\vert V_\mu^{{\rm lat}} \vert K\rangle Z_V=(m_s-m_q)
\langle \pi\vert S^{{\rm lat}} \vert K\rangle\,
\ea
with $Z_V$ a lattice renormalization factor for the vector
current. The scalar current is $S=\bar s q$, where field $q$ represents degenerate 
$u$ and $d$ quarks in our simulations. 
In this work, we use the local scalar density  of 
staggered fermions, as described in the following section, 
so the combination $(m_s-m_q)S^{{\rm lat}}$ requires no renormalization ($Z_S=1$). 
Using the definition of the form factors in Eq.~(\ref{eq:formfac})
and the identity in (\ref{eq:WI}), one can extract $f_0(q^2)$ at any 
value of the momentum transfer $q^2$ by using
\ba\label{eq:WIresult}
f_0(q^2) = \frac{m_s-m_q}{m_K^2-m_\pi^2}\langle \pi\vert S
\vert K\rangle_{q^2}.
\ea
Kinematic constraints demand that $f_+(0)=f_0(0)$, so this relation can be
used to calculate $f_+(0)$, and thus to extract $\vert V_{us}\vert$, 
the goal of this work. The main advantage of relation (\ref{eq:WIresult}) 
is that it does not need a renormalization factor to obtain the form factor $f_0(q^2)$. 
With staggered fermions, another key point is that it avoids the calculation of 
correlation functions with a vector current insertion. The staggered local vector 
current,  unlike the staggered local scalar current, is not a taste singlet. This 
implies that, in order to have a non-vanishing signal, we would need to use either 
external Goldstone mesons (with pseudoscalar taste) with a non-local vector 
current or external non-Goldstone mesons with a local vector current~\cite{HPQCD09}. 
Those correlation functions typically have larger statistical errors than the ones for 
a correlation function with a local current and external Goldstone mesons~\cite{Koponen:2012di}.

\begin{figure}[tb]
\begin{center}

\vspace*{-1.cm}
\includegraphics[width=0.65\textwidth]{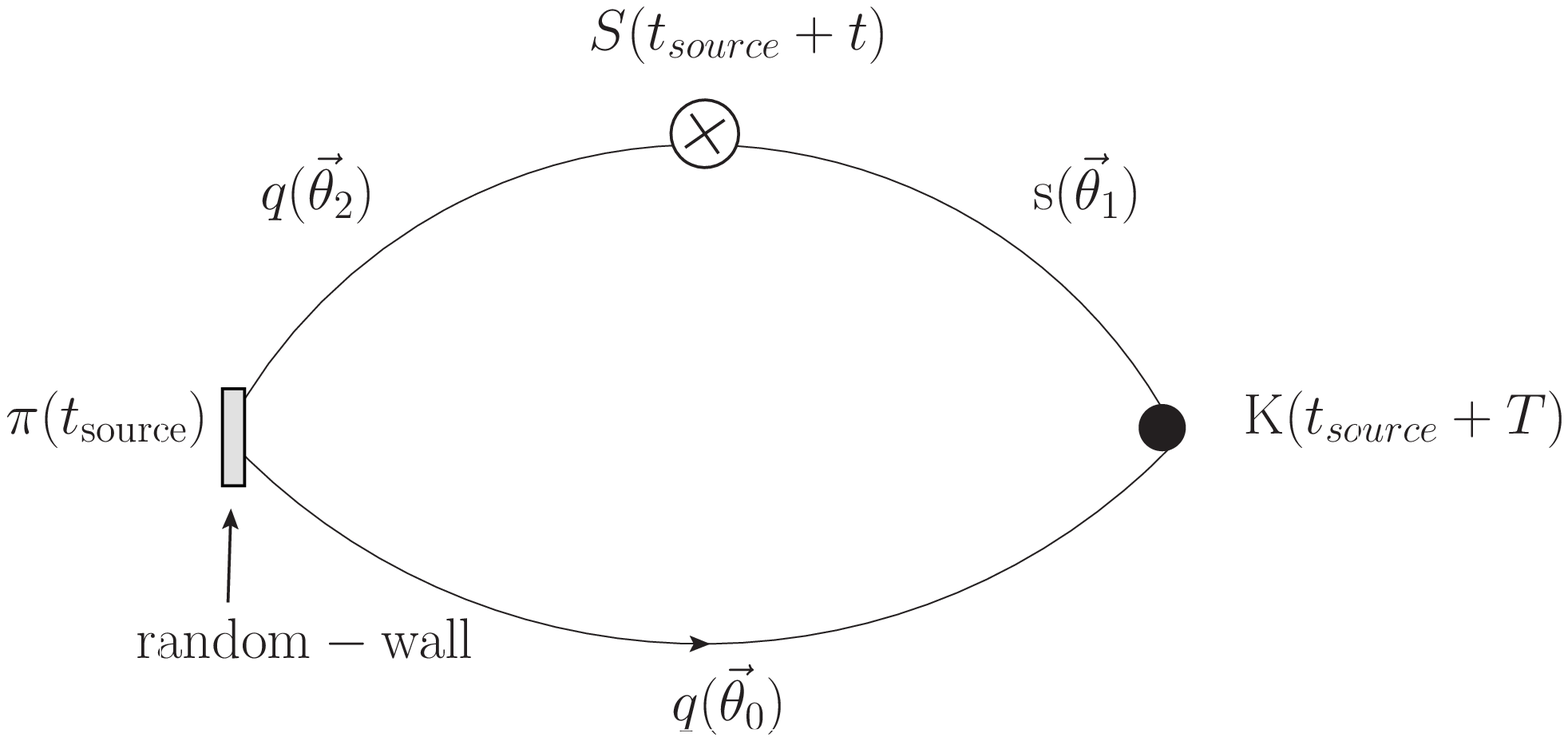}
\end{center}
\vspace*{-8.8cm}
\caption{Structure of the three-point functions needed to
calculate $f_{0}(q^2)$. Light-quark propagators are
generated at $t_{{\rm source}}$ with random-wall sources. An 
extended strange propagator is generated at $T+t_{{\rm source}}$. 
\label{fig:diagram}}
\end{figure}

The general structure of the three-point function that gives us access to the 
matrix element in Eq.~(\ref{eq:WIresult}) is depicted in Fig.~\ref{fig:diagram}. 
We generate light quarks at a time slice $t_{{\rm source}}$ and extended 
strange propagators at a fixed distance $T$ from the source. 
At $t_{{\rm source}}$ we use random-wall sources in a manner similar
to that of Ref.~\cite{McNeile:2006bz}. Specifically, we compute quark propagators
from each of four Gaussian-stochastic color vector fields with support on all
three color components and all spatial sites at that source time. This method greatly
reduces statistical errors, as discussed in Ref.~\cite{latt2010}. 
We place the scalar current 
at a time position $t_{{\rm source}} + t$ between the source and the sink,  
and contract the extended strange propagator with a light propagator. We then 
study the $t$ dependence to isolate the desired matrix element.

Another key ingredient in our calculation is the use of partially twisted boundary 
conditions~\cite{Bedaque:2004ax,tbc}. The valence quarks are generated
with twisted boundary conditions that satisfy the relation
\ba\label{tbcdef}
\psi(x_k+L) = e^{i\theta_k}\psi(x_k)\,,
\ea
where the subindex $k$ labels the spatial direction, $k=1,2,3$, and $L$ is the 
spatial extent of the lattice box in which we are simulating. The sea quarks,  
however, obey periodic boundary conditions. A meson with valence quarks generated 
with twisted boundary conditions as in Eq.~(\ref{tbcdef}), acquires a momentum 
$\vec{p}$ with components $p_k = \pi\Delta\theta_k/L$, where $\Delta\theta_k$ 
is the difference between the twisting angles of the two valence quarks. We can thus 
inject an arbitrary momentum to external mesons by tuning the twisting angles 
of their valence quarks. In particular, we can tune the external momentum 
to simulate directly at $q^2\approx0$, avoiding an interpolation in $q^2$ and 
the corresponding systematic uncertainty.

In our calculation, in order to get $q^2\simeq0$, we inject momentum 
into either the kaon or the pion. 
For a nonzero $\vec{p}_K$ we choose $\vec{\theta}_0=\vec{\theta}_2=0$,
$\vec{\theta}_1\ne\vec{0}$, and for a nonzero $\vec{p}_\pi$ we choose 
$\vec{\theta}_0=\vec{\theta}_1=\vec{0}$, $\vec{\theta}_2\ne\vec{0}$ 
(see Fig.~\ref{fig:diagram} for definitions of $\vec{\theta}_{0,1,2}$). 
The twisting angles are thus fixed according to 
\ba\label{twistangle}
\left\vert\vec{\theta}_1\right\vert_{q^2=0} =  \frac{L}{\pi}\, \sqrt{\left(
\frac{m_K^2+m_\pi^2}{2m_\pi}\right)^2-m_K^2}\,\, ,\quad
\left\vert\vec{\theta}_2\right\vert_{q^2=0} =  \frac{L}{\pi}\, \sqrt{\left(
\frac{m_K^2+m_\pi^2}{2m_K}\right)^2-m_\pi^2}\, ,
\ea
in each ensemble. The same choice of twisting angles was first 
explored for the calculation of $K\to\pi l\nu$ form factors in 
Ref.~\cite{Guadagnoli:2005be} (in the quenched approximation) and
Ref.~\cite{Boyle:2007wg}. 

\section{Numerical Simulations}
\label{sec:numerical}

\begin{table}[t]
    \centering
\caption{\label{tab:ensemblesa} 
Parameters of the ensembles analyzed in this work.
The second and third columns show the approximate lattice spacing and the volume.
$am_l$ and $am_h$ are the light and strange sea-quark masses, respectively.
$N_{{\rm conf}}$ is the number of configurations analyzed from each ensemble, and 
$am_l^{{\rm val}}$ and $am_s^{{\rm val}}$ are the light and strange valence-quark masses. 
$N_{\rm sources}$ is the number of values of $t_{\rm source}$ used on each configuration. 
$N_T$ is the number of source-sink separations $T$ considered. The $r_1/a$
values are obtained by fitting the calculated $r_1/a$ to a smooth
function~\cite{Allton96}, as explained in Ref.~\cite{MILCasqtad}. The $a\approx 0.12~{\rm fm}$ 
ensemble with volume $28^3\times 64$ is used solely for the study of finite volume effects.}
\begin{tabular}{lcccccccccc}
\hline\hline
       & $\approx a$ (fm) & $\left(\frac{L}{a}\right)^3
\times\frac{L_t}{a}$ & $am_l/am_h$ & $N_{{\rm conf}}$ & $am_s^{{\rm val}}$ 
& $am_l^{{\rm val}}$ & $N_{{\rm sources}}$ & $N_T$ & $r_1/a$ & $m_\pi L$ \\
\hline
coarse & $0.12$ & $20^3\times 64$ & $0.020/0.050$ & $2052$ & $0.0491$ & $0.02806$ & $4$ & $5$ & 2.65 & 6.22 \\
       &        & $20^3\times 64$ & $0.010/0.050$ & $2243$ & $0.0495$ & $0.01414$ & $4$ & $8$ & 2.62 & 4.48 \\
       &        & $28^3\times 64$ & $0.010/0.050$ & $275$ & $0.0495$ & $0.01414$ & $4$ & $4$ & 2.62 & 6.27 \\
       &        & $20^3\times 64$ & $0.007/0.050$ & $2109$ & $0.0491$ & $0.00980$ & $4$ & $5$ & 2.63 & 3.78 \\
       &        & $24^3\times 64$ & $0.005/0.050$ & $2098$ & $0.0489$ & $0.00670$ & $8$ & $5$ & 2.64 & 3.84\\
\hline
fine   & $0.09$ & $28^3 \times 96$ & $0.0124/0.031$ & $1996$ & $0.0337$ & $0.0080$ & $4$ & $5$ & 3.72 & 5.78 \\

        &       & $28^3 \times 96$ & $0.0062/0.031$ & $1930$ & $0.0336$ & $0.0160$ & $4$ & $5$ & 3.70 & 4.14 \\
\hline\hline
\end{tabular}
\end{table}

We have performed these calculations on the $N_f=2+1$ MILC 
ensembles~\cite{MILCasqtad,Bernard:2006nj,Bernard:2001av} with dynamical quarks 
simulated using the asqtad improved staggered action~\cite{lepage-1999-59}. 
The up and down quarks in these configurations are degenerate and heavier than the 
physical ones. The strange-quark mass is tuned to values near the physical one. The 
gluon action is a one-loop Symanzik improved and tadpole improved action, with 
errors of $O(\alpha_s a^2)$ because the one-loop correction from fermion loops was 
not included in the coefficients---see Ref.~\cite{MILCasqtad} and references 
therein. The configurations on the $N_f=2+1$ MILC ensembles were 
generated using the fourth-root procedure for eliminating extra degress of freedom 
(tastes) originating from fermion doubling. Despite nonlocality and unitariry 
violations at non-zero lattice spacing~\cite{Prelovsek05,BGS06}, there are 
numerical~\cite{Follana:2004sz,Durr:2003xs,DHW04,Donald:2011if} and 
theoretical~\cite{Bernard06,Bernard:2006vv,Shamir04,BGS08,Adams:2008db} arguments 
that indicate that the correct theory of QCD is recovered in the continuum limit.

For the valence fermions, we use the highly improved staggered quark (HISQ) 
action~\cite{Hisqaction}, which has much smaller taste-symmetry violations 
than the asqtad action. 
The mixed action setup, with a different light-quark formulation describing the 
sea and the valence quarks, increases the complexity of the analysis and could also 
increase some systematic errors, such as that from the chiral extrapolation. 
However, the effects of having a mixed action can be accounted for by using staggered 
chiral perturbation theory (S$\chi$PT), as described in Appendix~\ref{app:SCHPT}. 
Indeed, taste-breaking effects are the dominant source of discretization effects in 
our calculation, so reducing them, even just in the valence sector, is highly desirable.

We have performed simulations at two different values of the lattice spacing   
for the choice of parameters shown in Table~\ref{tab:ensemblesa}. 
The valence strange-quark mass is tuned to its physical value on each 
ensemble by matching $m_{\eta_s}$ to its correct value~\cite{r1HPQCD}~\footnote{The 
$\eta_s$ is a ficticious particle made of a strange quark-antiquark pair, considered 
without the disconnected contribution. Its ``physical value'' was determined 
in Ref.~\cite{r1HPQCD}.}. The valence 
light-quark masses are fixed according to the relation
\ba
\frac{m_l^{{\rm val}}({\rm HISQ})}{m_s^{{\rm phys.}}({\rm HISQ})} = 
\frac{m_l^{{\rm sea}}({\rm asqtad})}{m_s^{{\rm phys.}}({\rm asqtad})}\,.
\ea

We average results over four time sources separated by 16 (24) time slices on the
$a\approx0.12~{\rm fm}$ ($0.09~{\rm fm}$) ensembles but displaced by a random distance
from configuration to configuration to mitigate autocorrelations. The exception 
is the most chiral ensemble, in which we doubled the statistics and  
thus have time sources separated by 8 time slices. We increased the statistics on 
this ensemble because we found getting stable results for the correlator fits to be more 
challenging. 
In order to disentangle the desired matrix elements from the contamination 
due to oscillating states in the correlation functions, 
we consider several source-sink separations (time separation $T$ in 
Fig.~\ref{fig:diagram}), including both odd and even values, 
for each choice of quark masses and time source.  
We used the ensemble with $a\approx 0.12~{\rm fm}$ and sea quark masses $0.010/0.050$ 
to tune the optimal parameters for the simulations on the rest of the 
ensembles, so the number of source-sink separations analyzed is larger than 
for the other ensembles~\cite{latt2011}.

For each ensemble, we generate zero-momentum two-point $\pi$ and $K$ correlation functions, 
and two-point $\pi$ and $K$ correlation functions with momentum given by the twisting  
angles in Eq.~(\ref{twistangle}). The structure of these correlators is 

\ba
C_{{\rm 2pt}}^P(\vec{p};t) = \frac{1}{L^3}\sum_{\vec{x}}\sum_{\vec{y}} 
\,\langle\Phi_P^{\vec{p}}(\vec{y},t+t_{{\rm source}})\Phi_P^{\vec{p}\,\dagger}
(\vec{x},t_{\rm source})\rangle
\,,
\ea
where $\Phi_P^{\vec{p}}(\vec{x},t)$ is an interpolating operator creating a meson 
$P=\pi,K$ at time $t$ carrying momentum $\vec{p}$ generated by using twisted boundary 
conditions. The $L^{-3}\sum_{\vec{x}}$ is implemented via random wall sources.  
We also generate three sets of three-point functions with $q^2=q^2_{{\rm max}}$ (both pions 
and kaons at rest), $q^2\simeq 0$ with kaons at rest and moving pions, and $q^2\simeq 0$ with 
pions at rest and moving kaons. These correlators are obtained using 
\ba
C_{{\rm 3pt}}^{K\to\pi}(\vec{p_\pi},\vec{p}_K;t,t_{{\rm source}},T) =  
\frac{1}{L^3}\sum_{\vec{x}}\sum_{\vec{y}}
\sum_{\vec{z}} 
\,\langle \Phi_K^{\vec{p}_K}(\vec{x},t_{{\rm source}}+T)S(\vec{z},t)\Phi_\pi^{\vec{p}_\pi\,\dagger}
(\vec{y},t_{\rm source})\rangle\,\,
\ea
where either $\vec{p}_K$, $\vec{p}_\pi$, or both are zero. 
The scalar current is a taste-singlet current given by 
\ba 
S(\vec{z},t)=\bar\psi_s(\vec{z},t)\psi_q(\vec{z},t)\,,
\ea
where the $\psi$ fields are naive fields.  
 
The way in which the correlation functions above are expressed in terms of single-component 
staggered fields is explained in detail in Ref.~\cite{HPQCD_Dtopi}. 
The only difference between the correlation functions in that work and here 
is that we inject the external momenta using twisted boundary conditions.

\subsection{Fitting Method and Statistical Errors}

\label{sec:fitting}

We fit the two-point functions for a pseudoscalar meson $P$ to the expression
\ba\label{2pt}
C_{{\rm 2pt}}^{P} (\vec{p}_P;t) & = & \sum_{m=0}^{N_{exp}} (-1)^{m(t+1)}(Z_m^P)^2
\left(e^{-E_P^m t}+e^{-E_P^m(L_t-t)}\right)\,,
\ea
where $L_t$ is the temporal size of the lattice. Oscillating terms with 
$(-1)^{m(t+1)}$ do not appear for pions with zero momentum. In 
Table~\ref{tab:2ptfits} we summarize the masses of the valence (Goldstone) 
pions and kaons that determine the twisting angles and, thus, the external momentum 
$\vec{p}_P$  injected in the three-point functions to get $q^2=0$, together with 
these twisting angles and momenta for each ensemble. We also include in 
Table~\ref{tab:2ptfits} the sea Goldstone and root-mean-squared (RMS) 
pion masses from Ref.~\cite{Aubin:2004fs}. 

\begin{figure}[tb]
\includegraphics[angle=-90,width=0.65\textwidth]{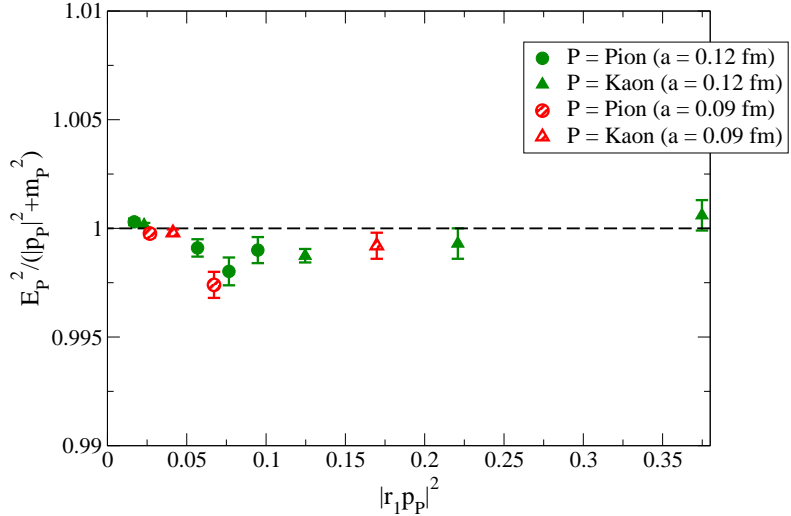}

\caption{Ratio of the measured (lattice) and the continuum dispersion relation for all 
data analyzed.   
\label{fig:dispersion}}
\end{figure}

From two-point function  
fits, we checked whether the continuum dispersion relation is satisfied. 
This is plotted in Fig.~\ref{fig:dispersion}, which shows very small deviations 
from the continuum prediction ($\le 0.25\%$), indicating small discretization 
effects. These deviations do not exceed power counting estimates of 
$\order\left(\alpha_s(a\vert\vec{p}\vert)^2\right)$ errors, except for the
point corresponding to a moving pion with $|r_1p_\pi|^2\approx 0.07$ and $a=0.09~{\rm fm}$, 
for which the power counting estimate is about 0.14\%. 
The data corresponding to a moving kaon on the same ensemble (the rightmost red 
(speckled) triangle in Fig.~\ref{fig:dispersion}), however, does not show any unusual behavior.   
In addition, the results for $f_+(0)$ as extracted from data with moving pions 
and moving kaons agree within one statistical $\sigma$ 
($\le 0.15\%$ error). This gives us confidence that the errors quoted below for 
$f_+(0)$ properly take into account the discretization effects seen in the dispersion 
relation in Fig.~\ref{fig:dispersion}.  

\begin{table}[t]
    \centering
\caption{\label{tab:2ptfits}
Sea Goldstone and RMS pion masses, valence Goldstone pion 
and kaon masses, twisting angles, and external momenta injected 
in the three-point functions for each ensemble. The quark masses $am_l$ and $am_h$  
are the same as in Table~\ref{tab:ensemblesa}, and 
the twisting angles $\vec{\theta}_1$ and $\vec{\theta}_2$ are defined in 
Fig.~\ref{fig:diagram} and Eq.~(\ref{twistangle}). The subscript $P$ in the meson 
masses refers to the pseudoscalar taste, as defined in Eq.~(\ref{eq:mesonmass}). 
The first line for each ensemble 
corresponds to moving pions and the second line to moving kaons.} 
\begin{tabular*}{\textwidth}{ @{\extracolsep{\fill}} ccccccccc}
\hline\hline
$\approx a$ (fm) & 
$am_l/am_h$ & $aM_{\pi,{\rm P}}^{sea}$ & $aM_{\pi,{\rm RMS}}^{{\rm sea}}$ &  
$aM_{\pi,{\rm P}}^{{\rm val}}$ & $aM_{K,{\rm P}}^{{\rm val}}$ 
& $\vert\vec{\theta}_1\vert$ & $\vert\vec{\theta}_2\vert$ & 
$\vert a\vec{p}_P\vert$  \\
\hline
$0.12$ &  
$0.020/0.050$ & 0.31125 & 0.3728  & 0.31315 & 0.36617 
& 0 & 0.31310 & 0.04918 \\ &
                  &           &  & & & 0.36610 & 0 & 0.05751 \\ & 
$0.010/0.050$ & 0.22447 & 0.3041  & 0.22587 & 0.33456 
& 0 & 0.57960 & 0.09104 \\ &
                 &   & & & & 0.85851  & 0 & 0.13485\\ & 
$0.007/0.050$ & 0.18891 & 0.2789  & 0.18907 & 0.32119 & 0 
& 0.66807 & 0.10494 \\ &
           &   & & & & 1.13486 & 0 & 0.17826\\ &
 $0.005/0.050$ & 0.15971 & 0.2600 & 0.15657 & 0.31225 & 0 
& 0.88804 & 0.11624 \\ &
           &   & & & & 1.76565 & 0 & 0.23112\\
\hline
$0.09$ & 
$0.0124/0.031$  & 0.20635 & 0.2247  & 0.20341 & 0.25241 & 0 
& 0.39257 & 0.04405 \\&
             &   & & & & 0.48667 & 0 & 0.05460\\ &
  $0.0062/0.031$ & 0.14789 & 0.1726 & 0.14572 & 0.23171 & 0 
& 0.62420 & 0.07003 \\ &
             &   & & & & 0.99253 & 0 & 0.11136\\
\hline\hline
\end{tabular*}
\end{table}

The functional form for the three-point functions is
\ba\label{3pt}
        C_{{\rm 3pt}}^{K\to \pi} (\vec{p}_\pi,\vec{p}_K;t,T) & = &
\sum_{m,n=0}^{N_{exp}^{3pt}} (-1)^{m(t+1)} (-1)^{n(T-t+1)}A^{mn}(q^2)Z_m^\pi Z_n^K\nonumber\\
&&\times\left(e^{-E_\pi^mt}+e^{-E_\pi^m(L_t-t)}\right)
\left(e^{-E_K^n(T-t)}+e^{-E_K^n(T-L_t+t)}\right)\,,
\ea
where the factors $Z_i^P$ are the amplitudes of the two-point functions 
in Eq.~(\ref{2pt}). The three-point parameter $A^{00}(q^2)$ in Eq.~(\ref{3pt}) is related to 
the desired form factor $f_0(q^2)$ via 
\ba 
f_0(q^2) = \frac{1}{2}A^{00}(q^2)\, \sqrt{2E_\pi E_K}\,
\frac{m_s-m_q}{m_K^2-m_\pi^2}\,,
\ea 
where we have used Eq.~(\ref{eq:WIresult}) and taken into 
account overall factors involved in the parametrization of the correlation 
function. We extract the form factors $f_0(q^2)$, using the expression above,  
directly from simultaneous Bayesian fits of the relevant three- and two-point functions.  
In these fits, we include several three-point functions with different values 
of the source-sink separation $T$, with at least one odd $T$ and one even $T$ to 
control the contributions from the oscillating states. We perform 
three sets of fits: including only the data corresponding to moving pions, 
including only the data corresponding to moving kaons, or including all data. 
In the last case, we assume that the parameters $A^{mn}(q^2)$ depend only on $q^2$, 
so they are the same for moving pions or moving kaons. This is 
true up to discretization errors of the order of the deviations from the continuum 
dispersion relation of those data ($\le 0.25\%$). The quality of all the fits included in 
our determination of the form factor is rather high, with p-values ranging between 
$0.6$ and $1$. This may indicate that there is some mild overestimation of statistical errors. 

\begin{figure}[tbh]
\begin{minipage}[c]{.45\textwidth}
\hspace*{-0.4cm}\includegraphics[angle=-90,width=1.1\textwidth]{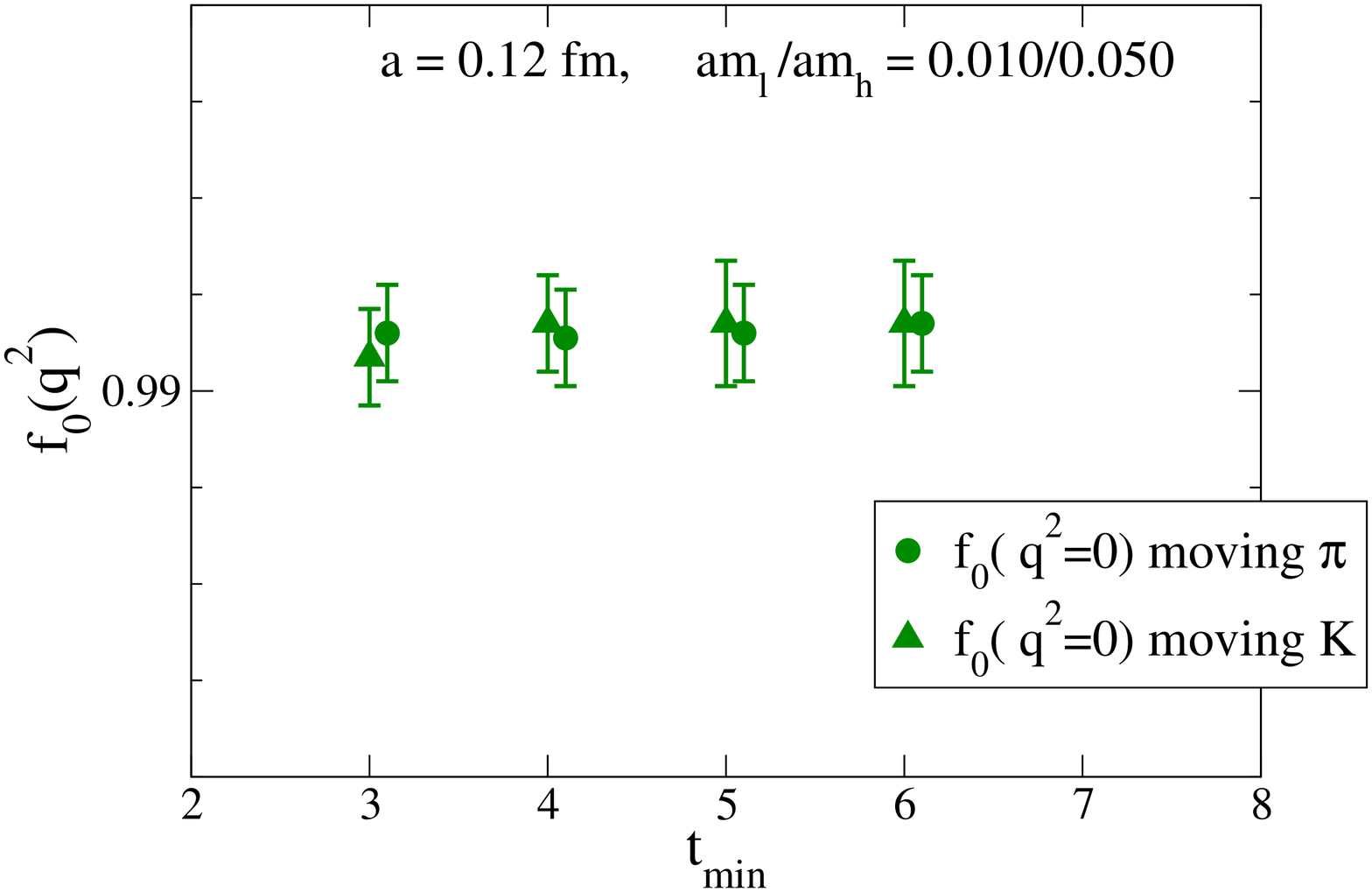}
\hspace*{-0.4cm}\includegraphics[angle=-90,width=1.1\textwidth]{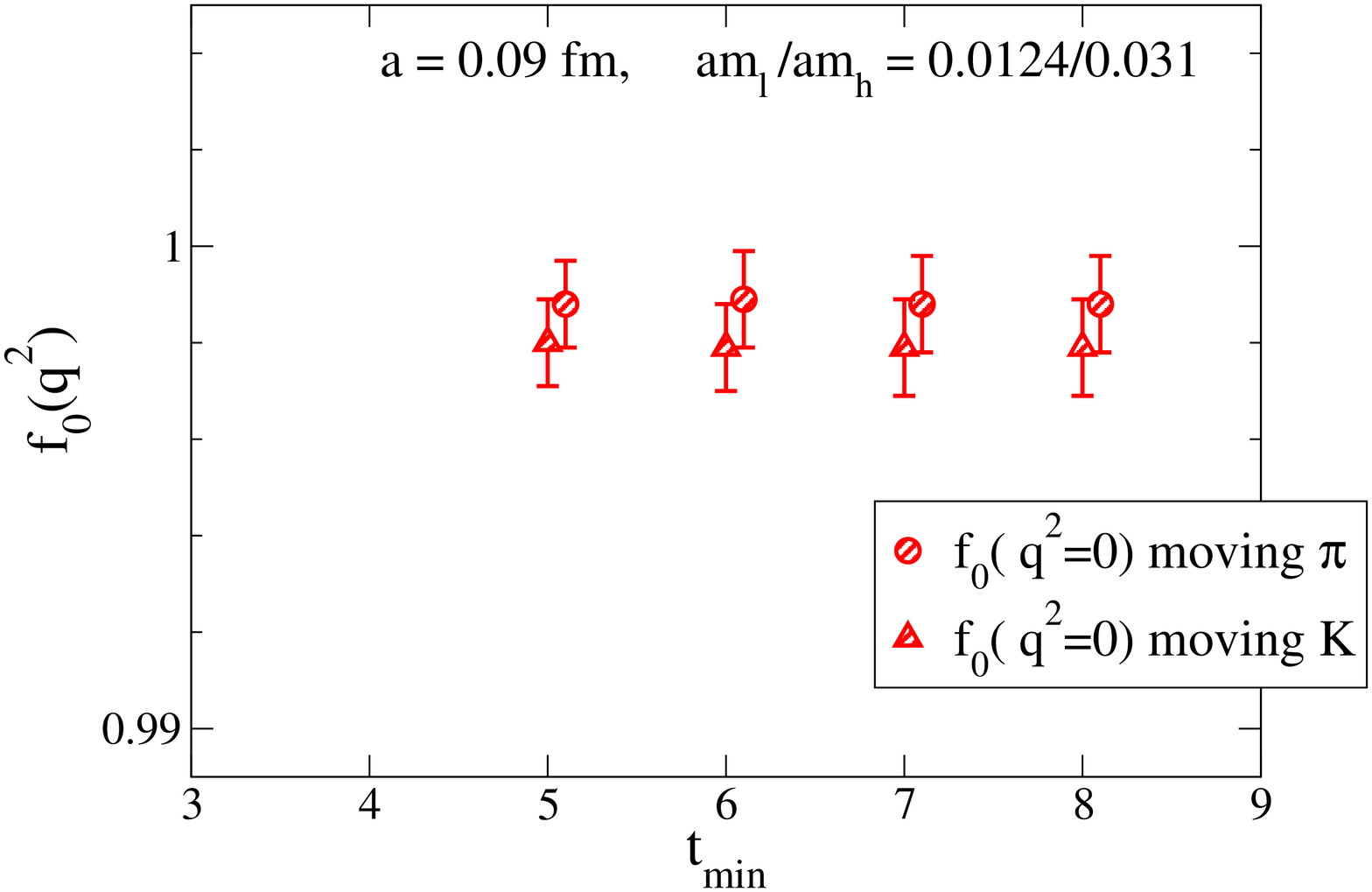}
\end{minipage}
\begin{minipage}[c]{.45\textwidth}
\hspace*{0.8cm}\includegraphics[angle=-90,width=1.1\textwidth]{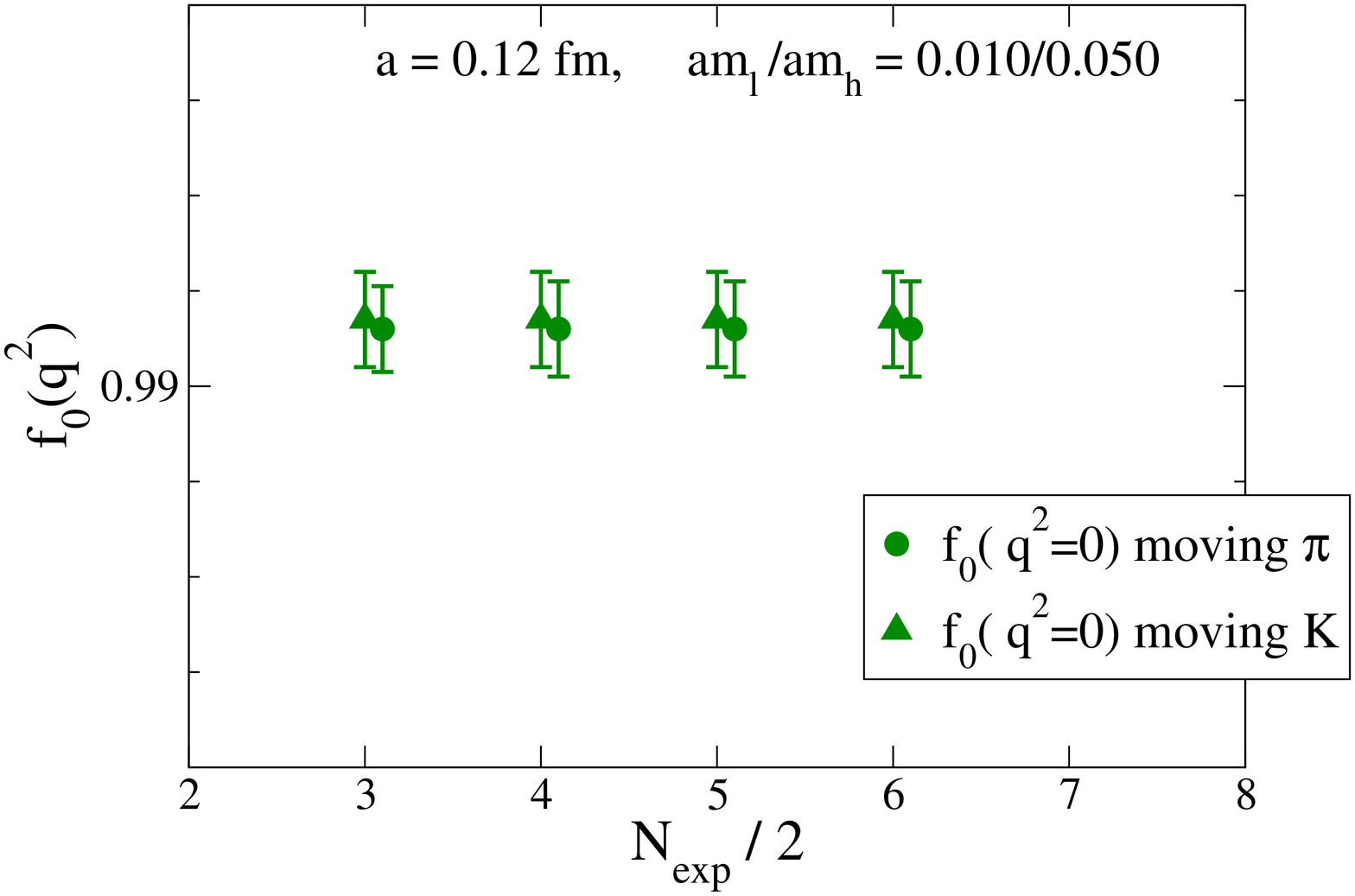}
\hspace*{0.8cm}\includegraphics[angle=-90,width=1.1\textwidth]{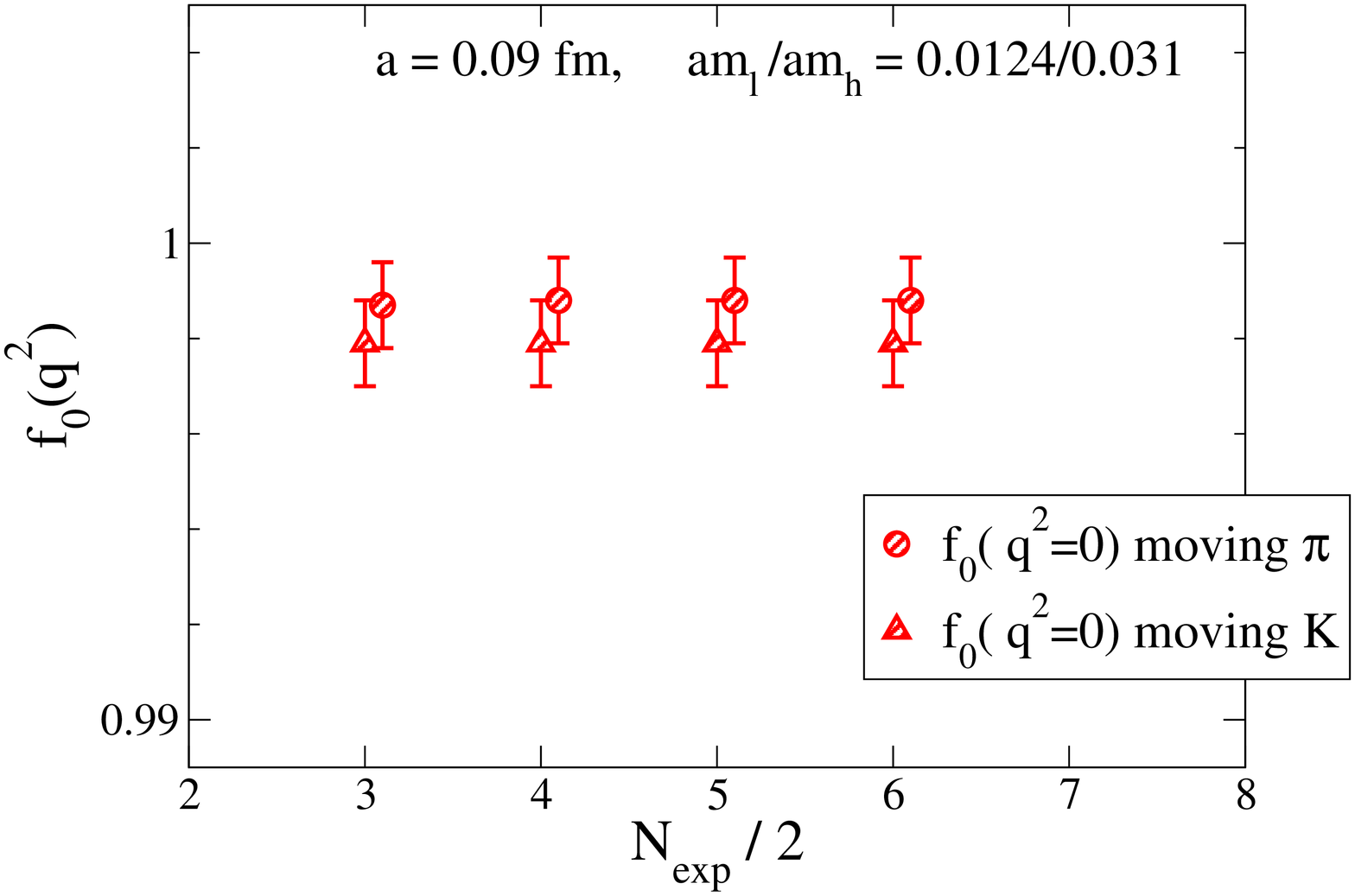}
\end{minipage}
\caption{Variation of $f_0(q^2)$ with $t_{{\rm min}}$ (left panel) and
$N_{{\rm exp}}$ (right panel) for $q^2=0$ generated by injecting momentum in
the $\pi$ or the $K$, and for $q^2=q_{{\rm max}}^2$ (zero external momentum) 
for two of the ensembles analyzed. 
\label{fig:stability} }
\end{figure}

In this analysis, it is especially relevant to check for the stability of our fits 
under the choice of fitting parameters and techniques, since we have very small 
statistical errors and need to be sure that these results are not dependent  
on our methodology. One check is to vary the time fitting 
ranges and number of states included in the fits. The fitting range is  
$t\in\left[t_{{\rm min}},(L_t-t_{{\rm min}})\right]$ for two-point functions and 
$t\in\left[t_{{\rm min}},(T-t_{{\rm min}})\right]$ for three-point functions---see 
Fig.~\ref{fig:diagram} and Table~\ref{tab:ensemblesa} for notation. The number of states 
included is the same in the regular and oscillating sectors, so
$N_{{\rm exp}}/2=N_{{\rm regular\,states}}=N_{{\rm oscillating\,states}}$. 
We also keep the number of states the same in all 
correlation functions included in a given simultaneous fit. Fixing $N_{{\rm exp}}$ 
and changing $t_{{\rm min}}$ from 3 (5) for $a\approx0.12~{\rm fm}$ ($0.09~{\rm fm}$) 
ensembles up to the maximum allowed by the source-sink separation gives us a 
plateau for the central  values with only small variations in the errors. Analogously, 
fixing $t_{{\rm min}}$ to our preferred value, we do not find any significant variation 
of the results for $N_{{\rm exp}}\ge 6$--$8$. Examples of these variation tests for 
two of the ensembles analyzed are shown in Fig.~\ref{fig:stability}.

We also study which combination of $T$'s is optimal. 
We find that the central values are very insensitive to the number of three-point 
functions included and the values of $T$ in the range we are analyzing. 
Errors and stability are better when $ 15\le T \le 24$ with three 
values of $T$ in the three-point functions for the $a\approx0.12~{\rm fm}$ ensembles, 
and $ 18\le T \le 33$ with four values of $T$ in the three-point 
functions for the $a\approx0.09~{\rm fm}$ ensembles. 

Finally, we check an alternative fit procedure, using the 
superaverage method described in Ref.~\cite{Btopi}. This takes an explicit
combination of three-point functions with consecutive values of $T$ and the time 
slice $t$ that suppresses the contribution from both the first regular 
excited state and the first oscillating state: 
\ba 
\label{3ptISA}
        \bar{C}_{3pt}^{K\to \pi} (t,T) & = &
 \frac{e^{-E_\pi^{(0)}t}
\, e^{-E_K^{(0)}(T-t)}}{8} \nonumber\\
                & \times & \bigg[ \frac{C_{3pt}^{K\to\pi} (t,T)}
{e^{-E_\pi^{(0)}t} e^{-E_K^{(0)}(T-t)}} + \frac{C_{3pt}^{K\to\pi}
(t,T+1)}{e^{-E_\pi^{(0)}(t)} e^{-E_K^{(0)}(T+1-t)}}  +  \frac{2 \,
C_{3pt}^{K\to\pi} (t+1,T)}{e^{-E_\pi^{(0)}(t+1)} e^{-E_K^{(0)}(T-t-1)}}
\nonumber\\
                & + & \frac{2 \, C_{3pt}^{K\to\pi} (t+1,T+1)}
{e^{-E_\pi^{(0)}(t+1)} e^{-E_K^{(0)}(T-t)}} +  \frac{C_{3pt}^{K\to\pi}
(t+2,T)}{e^{-E_\pi^{(0)}(t+2)} e^{-E_K^{(0)}(T-t-2)}} +
\frac{C_{3pt}^{K\to\pi} (t+2,T+1)}{e^{-E_\pi^{(0)}(t+2)}
 e^{-E_K^{(0)}(T-t-1)}} \bigg] \nonumber\\
                & \approx &  A^{00}\sqrt{Z_0^K Z_0^\pi}e^{-E_\pi^{(0)}t} \,
e^{-E_K^{(0)}(T-t)} \nonumber\\
&+& (-1)^{T} \bar{A}^{11} e^{- E_\pi^{(1)}t}
e^{-E_K^{(1)}(T-t)} +  \bar{A}^{20}
e^{-E_K^{(2)}(T-t)} e^{-E_\pi^{(0)}t} +  \bar{A}^{02} e^{- E_\pi^{(2)}t}
e^{-E_K^{(0)}(T-t)}\nonumber\\
& + & \order(\Delta E_\pi^2,\, \Delta E_\pi \Delta E_K,\, \Delta E_K^2)\,.
\ea
Again, results are compatible with our preferred fitting method within errors.

We also checked for autocorrelations in our data and did not observe any effect 
on the relevant correlation functions.  Central values and errors of 
the ground-state parameters remain unchanged when blocking by 2, 3, and 4. 
We thus do not perform any blocking of our data.

\begin{center}
\begin{figure}[tbh]
\includegraphics[angle=-90,width=0.65\textwidth]{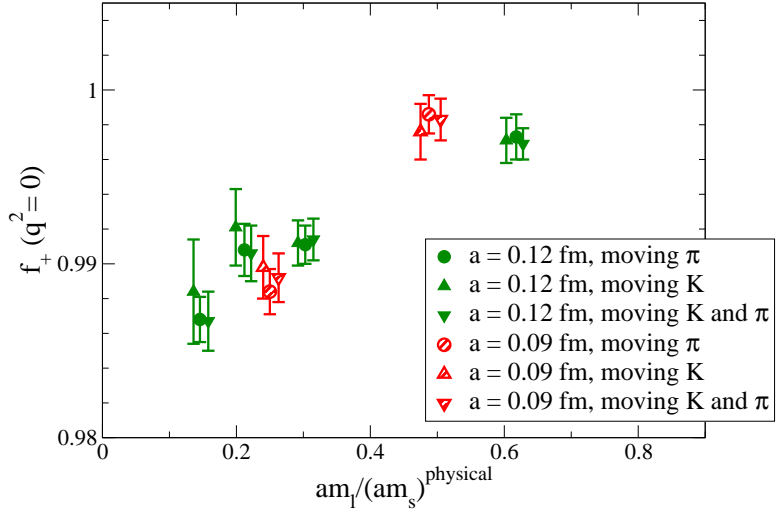}
\caption{Form factor $f_+(0)$ obtained from the different ensembles in
Table~I as a function of the valence light-quark mass normalized to the physical
strange-quark mass. The squares, upward-pointing triangles, and downward-pointing 
triangles correspond to fits to data with only moving pions, only moving kaons, and both
moving pions and moving kaons, respectively. The filled green (speckled red) 
symbols label ensembles with
$a\approx 0.12~{\rm fm}$ $(0.09~{\rm fm})$. We give a horizontal offset to two of the three
points at each light-quark mass for clarity.
\label{fig:data}}
\end{figure}
\end{center}

In Fig.~\ref{fig:data}, we collect the results for $f_+(0)$ from our preferred 
correlator fit methodology as a function of $am_l/am_s^{{\rm phys}}$. 
The errors shown are statistical only and generated with a 500-bootstrap distribution. 
They are very small, around $0.1$--$0.15\%$, except for the cases noted below. 
We plot results coming from only three-point functions where the external momentum to 
obtain $q^2\approx0$ is injected via the moving $\pi$ (circles), via the moving $K$ 
(upward-pointing triangles), and combined fits including both correlation functions with 
moving $\pi$ and moving $K$ (downward-pointing triangles). 
The correlator fits for the $0.005/0.050$ and $0.007/0.050$ ensembles with
$a\approx0.12~{\rm fm}$ and moving kaons are not as stable as for the other 
ensembles, and errors are considerably larger. These results correspond to the two first 
filled green upward-pointing triangles from the left in Fig.~\ref{fig:data}. 
This is due to the fact that  
we need a larger momentum in lattice units to 
get $q^2=0$ when reducing the light-quark masses, and correlators become noisier. For 
this reason, we drop these two sets of data from our analysis.

The results from fits of three-point functions with moving $K$, moving $\pi$, or 
with both sets of correlators  all 
agree within our very small statistical errors, as can be seen in Fig.~\ref{fig:data}.
This constitutes a strong test of our methodology and quoted errors. 
It also suggests that the discretization errors that break the continuum dispersion
relation (which contribute differently for $f_+(0)$ as extracted from moving pions and
moving kaons) are smaller than our statistical errors.

For the rest of the analysis, we take separately the data with moving pions and
moving kaons, omitting the two ensembles with moving kaons mentioned above. 
This choice allows us to identify more clearly the different sources of discretization 
errors. The numerical value of the data included in this analysis are listed in
Table~\ref{numericalfplus}.

\begin{table}[t]
    \centering
\caption{\label{numericalfplus}Values of $f_+(0)$ included in the chiral and continuum extrapolation. 
Errors are statistical only, from 500 bootstrap ensembles.} 
\begin{tabular}{lcccc}
\hline\hline
$\approx a$ (fm) & $am_l/am_h$ & $aM_{\pi,{\rm P}}^{{\rm val}}$ & $f_+^{{\rm moving\,\pi}}(0)$ 
& $f_+^{{\rm moving\,K}}(0)$\\
\hline
$0.12$ & $0.020/0.050$ & 0.31315 & 0.9973(13) & 0.9971(13) \\
& $0.010/0.050$ & 0.22587 & 0.9911(11) & 0.9912(13) \\
& $0.007/0.050$ & 0.18907 & 0.9908(15) & - \\ 
& $0.005/0.050$ & 0.15657 & 0.9868(13) & -\\ 
\hline
$0.09$ & $0.0124/0.031$  & 0.20341 & 0.9986(11) & 0.9976(16) \\
&  $0.0062/0.031$ & 0.14572 & 0.9884(11) & 0.9898(18) \\
\hline\hline
\end{tabular}
\end{table}

\section{Extrapolation to the physical point}

\label{sec:chiralfits}

In this section, we describe the use of chiral perturbation theory ($\chi$PT) to 
extrapolate our data at several values of the light quark masses down to the physical 
point. The form factor $f_+(0)$ can be written as a $\chi$PT expansion in the following way
\ba
f_+(0) = 1 + f_2 + f_4 + f_6 + ... = 1 + f_2 + \Delta f\,,
\ea 
where the $f_{2i}$'s contain corrections of $\order(p^{2i})$ in the chiral 
power counting. The Ademollo-Gatto (AG) theorem~\cite{AGtheorem}, 
which follows from vector current conservation, 
ensures that $f_+(0) \to 1$ in the SU(3) limit and, further, that the SU(3)  
breaking effects are second order in $(m_K^2-m_\pi^2)$. This fixes $f_2$ completely 
in terms of experimentally measurable quantities. At finite lattice spacing, however, we have 
violations of the AG theorem due to symmetry-breaking discretization effects
in the form-factor decomposition, Eq.~(\ref{eq:formfac}), and 
in the continuum dispersion relation needed to derive the relation between $f_0(0)$
and the correlation functions, Eq.~(\ref{eq:WIresult}). 
These and other discretization effects are very small in our data, 
though, as can be deduced from Fig.~\ref{fig:data}. 

We study the light-quark mass dependence and the discretization effects 
in our calculation using two-loop continuum $\chi$PT \cite{BT03}, 
supplemented by partially quenched staggered $\chi$PT (PQS$\chi$PT) at one loop. 
The small variation with $a$ in our data suggests that addressing those effects at 
one loop should be enough for our target precision. We also incorporate  
partial quenching effects in the continuum expression at 
one-loop~\cite{Bijnenspc,Becirevic:2005py}. The staggered one-loop fitting 
function can be found in Appendix~\ref{app:SCHPT}, while the details of its calculation 
will be given in Ref.~\cite{KtopilnuSChPT}. 

In order to test the dependence 
of our results on the fit function, as well as the systematics associated 
with disregarding higher order terms in both the chiral and the $a^2$ expansions, 
we also use another approach for the chiral and continuum extrapolation. 
We replace the continuum two-loop $\chi$PT with a general 
analytic NNLO parametrization. 
The implementation and results obtained with each of these methods are discussed 
in the following sections.

\subsection{Partially quenched S$\chi$PT at NLO plus continuum NNLO  
$\chi$PT}

\label{sec:chpttwooloop}

We take the NLO staggered partially quenched $\chi$PT (PQS$\chi$PT) expression 
in Appendix~\ref{app:SCHPT} and add the two-loop full QCD continuum calculation from  
Ref.~\cite{BT03}. The one-loop corrections that relate the pion decay constant $f_\pi$ 
to the SU(3) chiral limit, $f_0$, must be accounted for consistently when $f_\pi$ 
appears in the one-loop expression and we are working through two-loop order.  Our 
two-loop expression is chosen so that $f_\pi$ evaluated at the physical pion mass is 
the appropriate value to use in the coefficient of the one-loop chiral logarithms. 
We also add a term that parametrizes discretization effects 
arising from the violation of the continuum dispersion relation, which breaks the AG 
theorem, and another term of $\order(a^2)$ that respects the AG theorem. They are proportional 
to $K_1^{(a)}$ and $K_2^{(a)}$ respectively in Eq.~(\ref{eq:ChPTtwoloop}). This gives 
us the general fit function
\ba\label{eq:ChPTtwoloop}
f_+(0) = 1 + f_2^{{\rm PQS\chi PT}}(a) &+& K_1^{(a)}\,\left(\frac{a}{r_1}\right)^2 +
f_4^{{\rm cont.}}({\rm logs}) + f_4^{{\rm cont.}}(L_i)\nonumber\\
&+& r_1^4\,(m_\pi^2-m_K^2)^2 \left[C_6'^{(1)}
+ K_2^{(a)}\,\left(\frac{a}{r_1}\right)^2\right]\,,
\ea
where the constants $K_1^{(a)}$, $K_2^{(a)}$, and $C_6'^{(1)}\propto C_{12} + C_{34} 
- L_5^2$ are constrained parameters to be determined by the chiral fits using 
Bayesian techniques. All dimensionful quantities entering in the chiral fit  
functions are converted to $r_1$ units using the values of $r_1/a$ in 
Table~\ref{tab:ensemblesa}. We 
split the two-loop contribution, $f_4$, into a piece containing the logarithmic 
terms alone and another one depending on the $\order(p^4)$ low-energy constants (LEC's) 
$L_i$. The first piece does not contain any free parameters. For the 
NLO LEC's, we can either choose to fix them or leave them as constrained parameters in the 
fits. This choice does not make much difference in our final results, though, as explained 
in Sec.~\ref{sec:CHPTresults}. The $C_{ij}$ are $\order(p^6)$ LEC's defined in 
Ref.~\cite{ChPTp6}. Notice that only one combination of $\order(p^6)$ LEC's contributes to
the chiral expansion for $f_+(0)$ at this order, so only one extra free parameter
enters at NNLO.

The NLO  partially quenched $\chi$PT (PQS$\chi$PT) function, $f_2^{{\rm PQS\chi PT}}(a)$,  
incorporates the dominant lattice artifacts from taste-symmetry breaking. The only unknown 
quantities entering in this function are the ``mixed'' taste-violating hairpin 
parameters, $\delta_V^{{\rm mix}}$ and $\delta_A^{{\rm mix}}$, defined in Appendix~\ref{app:SCHPT}. 
These coefficients appear because we use a mixed action, and we expect their values  
to be small. Thus, we take them as constrained parameters in the fit, with prior 
central values equal to zero and prior widths equal or larger than the hairpin 
parameters in the valence sector, $\delta_V^{{\rm HISQ}}$ and $\delta_A^{{\rm HISQ}}$. 
The remaining inputs needed to obtain $f_2^{{\rm PQS\chi PT}}(a)$ have been 
calculated by our collaboration or can be taken from experiment. Their values 
are collected in Table~\ref{tab:CHPTparameters},   
and their definitions appear in Appendix~\ref{app:SCHPT}. The asqtad 
taste splittings are taken from Ref.~\cite{Aubin:2004fs}, 
while the HISQ splittings are preliminary results from the MILC Collaboration that 
agree well with their final results in Ref.~\cite{hisqconfig}. 
We extract the mixed taste splittings from simulations on the $a\approx0.12~{\rm fm}$ 
ensemble with sea quark masses $0.010/0.050$ and valence light-quark mass $am_l=0.2am_s$. 
With the exception of the pseudoscalar taste splitting, which is non-zero for a mixed action,
the other mixed taste splittings agree fairly well with the values given by the
approximation used in Ref.~\cite{latt2012}:
$\Delta_\Xi^{{\rm mix}}=(\Delta_\Xi^{{\rm asqtad}}+\Delta_\Xi^{{\rm HISQ}})/2$. However, we 
observe a systematic effect that the real taste splittings are between 10\% and 30\% larger than 
those from this approximation. Since the fits for the tensor taste are not stable enough to 
extract the corresponding taste splitting, we use this effect as a guide 
and estimate $\Delta_T^{{\rm mix}}$ to be $20\%$ larger than the sea-valence average. 
The final result is insensitive to this choice, however.  
Finally, the taste-violating hairpin parameters for the HISQ action, collected 
also in Table~\ref{tab:CHPTparameters}, are taken from chiral fits to the HISQ light 
pseudoscalar data \cite{Bazavov:2011fh}.

\begin{table}[t]
    \centering
\caption{Inputs for the fixed parameters needed in the NLO PQS$\chi$PT 
part of the fit functions in Eqs.~(\ref{eq:ChPTtwoloop}) and 
Eqs.~(\ref{eq:ChPTfunction}). We do not consider errors on the slope $\mu r_1$, 
the pion decay constant $f_{\pi}r_1$, or the taste splittings $r_1^2a^2\Delta_\Xi$ because 
they have negligible effect on the final results. \label{tab:CHPTparameters}}
\begin{tabular}{ccc}
\hline\hline
 & \hspace*{-0.3cm}$a\approx 0.12~{\rm fm}$ & $a\approx 0.09~{\rm fm}$ \\
\hline
$f_{\pi}r_1$ & \multicolumn{2}{c}{0.20725}\\
$\Lambda_{\chi} r_1=M_\rho r_1$ & \multicolumn{2}{c}{1.2225}\\  
\hline
$\mu r_1$ (asqtad) & 6.234 & 6.382  \\
\hline
$r_1^2a^2\Delta_P^{{\rm asqtad}}$ & 0 & 0 \\
$r_1^2a^2\Delta_V^{{\rm asqtad}}$ & 0.439 & 0.152 \\
$r_1^2a^2\Delta_T^{{\rm asqtad}}$ & 0.327 & 0.115  \\
$r_1^2a^2\Delta_A^{{\rm asqtad}}$ & 0.205 & 0.0706  \\
$r_1^2a^2\Delta_I^{{\rm asqtad}}$  & 0.537 & 0.206 \\
\hline
$r_1^2a^2\Delta_P^{{\rm HISQ}}$ & 0 & 0 \\
$r_1^2a^2\Delta_V^{{\rm HISQ}}$ & 0.171 & 0.0573 \\
$r_1^2a^2\Delta_T^{{\rm HISQ}}$ & 0.112 & 0.0390 \\
$r_1^2a^2\Delta_A^{{\rm HISQ}}$ & 0.0575 & 0.0204 \\
$r_1^2a^2\Delta_I^{{\rm HISQ}}$  & 0.222 & 0.0709 \\
\hline
$r_1^2a^2\Delta_P^{{\rm mixed}}$ & 0.0828 & 0.0828$\times$0.35 \\
$r_1^2a^2\Delta_V^{{\rm mixed}}$ & 0.325 & 0.325$\times$0.35  \\
$r_1^2a^2\Delta_T^{{\rm mixed}}$ & 0.263 & 0.263$\times$0.35  \\
$r_1^2a^2\Delta_A^{{\rm mixed}}$ & 0.174 & 0.174$\times$0.35  \\
$r_1^2a^2\Delta_I^{{\rm mixed}}$ & 0.403 & 0.403$\times$0.35 \\
\hline\hline
\end{tabular}
\end{table}

\subsection{Partially quenched S$\chi$PT at NLO plus NNLO analytical parametrization}

To estimate systematic errors, we use a second approach. We 
take the same NLO staggered partially quenched $\chi$PT 
expressions as before, but, instead of the two-loop continuum calculation, 
we add a general analytic parametrization of higher order terms and $a^2$ corrections of 
the form 
\ba \label{eq:ChPTfunction}
f_+(0) = 1 + f_2^{PQS\chi PT}(a) &+& K_1^{(a)} \left(\frac{a}{r_1}\right)^2 
+ r_1^4(m_\pi^2-m_K^2)^2 \left\lbrack C_6^{(1)} + C_8^{(1)}(r_1 m_\pi)^2
\vphantom{\left(\frac{a}{r_1}\right)^2}\right.\nonumber\\
&+& \left.C_8^{(2)}(r_1 m_\pi)^2\ln (m_\pi^2/\Lambda_\chi^2) + C_{10}^{(1)} (r_1 m_\pi)^4 +
K_2^{(a)}\left(\frac{a}{r_1}\right)^2\right\rbrack,\,
\ea
where $\Lambda_\chi$ is the chiral scale, which we take equal to the mass of the $\rho$ 
throughout this analysis. The discretization effects in Eq.~(\ref{eq:ChPTfunction}) 
are parametrized in the same way as in the 
fit function in Eq.~(\ref{eq:ChPTtwoloop}). The terms proportional to 
the coefficients $C_{2j}^{(i)}$ are of $\order(p^{2j})$ in the chiral expansion, 
so $C_6^{(1)}$ is NNLO as before, $C_8^{(i)}$ are NNNLO, and $C_{10}^{(1)}$ 
is NNNNLO. Notice that we do not include a term 
of order $(r_1 m_K)^2$ in the expression above since the valence strange-quark masses 
are tuned to the physical value and are thus the same in units of $r_1$. 
The effect produced by the difference 
between the strange-quark mass in the valence and the sea sectors is discussed 
in Sec.~\ref{sec:fitfunction}.

  \begin{table}[tb]
    \centering
\caption{
Priors for the fit parameters entering the expressions in 
Eqs.~(\ref{eq:ChPTtwoloop}) and~(\ref{eq:ChPTfunction}). 
The dimensionless $\chi$PT parameter $s$ is given by the quantity $1/(8\pi^2(r_1f_\pi)^2)$.
\label{tab:priors} See the text for explanation of the choice of priors. The priors 
listed for the hairpin parameters are for the $a\approx0.12~{\rm fm}$ 
ensembles, and those for the $a\approx0.09~{\rm fm}$ ensembles are obtained 
by multiplying by 0.35. This factor comes from assuming 
that the hairpin parameters scale like the $\Delta_\Xi$. 
The central values for the NLO LEC's are from fit 
10 in Ref.~\cite{LisHans}.}

\begin{center}  
\begin{tabular}{cc}
\hline
\hline
\multicolumn{2}{c}{Fit parameters (central value$\pm$width)} \\
\hline
\hline
$r_1^2a^2\delta_V^{{\rm HISQ}}$ & $0.057\pm0.033$\\
$r_1^2a^2\delta_A^{{\rm HISQ}}$ & $-0.0782\pm0.0040$\\
\hline
$r_1^2a^2\delta_V^{{\rm mix}}$ & $0.0 \pm 0.1$\\
$r_1^2a^2\delta_A^{{\rm mix}}$ & $0.0 \pm 0.1$\\
\hline
$K_{1,2}^{(a)}$ & $0\pm 1$ \\
$C_6'^{(1)}$ & $0\pm s^2$ \\
$C_6^{(1)}$ & $0\pm s^2$ \\
$C_8^{(i)}$ & $0\pm s^3$ \\
$C_{10}^{(1)}$ & $0\pm s^4$ \\
\hline
\hline
$L_1^r(M_\rho)\times 10^3$ &  $0.43\pm0.24$ \\ 
$L_2^r(M_\rho)\times 10^3$ &  $0.73\pm0.24$  \\
$L_3^r(M_\rho)\times 10^3$ &  $-2.30\pm0.74$ \\
$L_4^r(M_\rho)\times 10^3$ &  $0.0\pm0.6$   \\
$L_5^r(M_\rho)\times 10^3$ &  $0.97\pm0.22$     \\
$(2L_6^r-L_4^r)(M_\rho)\times 10^3$ &  $0.0\pm0.4$ \\
$L_7^r(M_\rho)\times 10^3$ &  $-0.31\pm0.28$   \\
$L_8^r(M_\rho)\times 10^3$ &  $0.60\pm0.36$     \\
\hline\hline
\end{tabular}
\end{center}
\end{table}

\subsection{Results}

\label{sec:CHPTresults}

We obtain the central value for $f_+(0)$ by extrapolating our data to 
the physical meson masses and the continuum limit using the fitting functional 
form in Eq.~(\ref{eq:ChPTtwoloop}) with 
$K_2^{(a)}=0$ and allowing for a non-zero value of $K_1^{(a)}$. 
We use Bayesian techniques and estimate the statistical errors by fitting to a set 
of 500 bootstrap samples for each ensemble. In these fits, we include results coming 
from only injecting external momentum in the $\pi$ and only injecting external 
momentum in the $K$, circles and upward-pointing triangles in Fig.~\ref{fig:data}, respectively. 
We omit the data with moving kaons in the two most chiral $a\approx0.12~{\rm fm}$ 
ensembles, due to the lower quality of those data, as explained in Sec.~\ref{sec:fitting}. 
For the meson masses we use the PDG values, $m_{\pi^-}=139.570~{\rm MeV}$ and  
$m_{K^0}=497.614~{\rm MeV}$.

The prior central values and widths we use in our fits for the constrained parameters are
summarized in Table~\ref{tab:priors}. For the priors for $C_6'^{(1)}$, we
assume that the NNLO analytical terms are of the same order as the NNLO logarithmic 
terms, which are $\sim s^2(r_1m_\pi)^4$ with $s$ 
a typical $\chi$PT scale, $s \equiv 1/(8\pi^2(r_1f_\pi)^2)$. Following the same 
reasoning, we choose the priors widths to be $s^3$ and $s^4$ for the coefficients 
of the NNNLO and NNNNLO terms, respectively. We set the priors for the coefficients of 
the $a^2$ terms to $0\pm1$ and
check that the output of the fits for those coefficients is not constrained by
that choice. For example, in our main fit the output for $K_1^{(a)}$ is 
$-0.011(9)$, well under the limits set by the corresponding prior. 

We treat the HISQ hairpin parameters as constrained parameters in our fits, with prior 
central values and widths equal to the values and errors obtained in 
Ref.~\cite{Bazavov:2011fh} from fits to other light quantities. Those priors are also 
listed in Table~\ref{tab:priors}. In this way the uncertainty associated with the 
error of those parameters is included 
directly in the statistical error for the extrapolated form factor. 
The same strategy is adopted for the NLO LEC's needed in the fit function,  
$L_i$ with $i=1$--$5$. We constrain their values using the results from 
fit 10 in Ref.~\cite{LisHans}~\footnote{The authors of Ref.~\cite{LisHans} fit experimental 
data on $K_{l4}$ decays to the corresponding two-loop $\chi$PT expressions to 
extract the value of the NLO LEC's. In fit 10, $L_4$ and $L_6$ are set to zero. More details 
on the type of fit performed can be found in Ref.~\cite{Amoros:1999qq}.} 
for the prior central values and 2 times the 
error in the same reference for the prior widths; see Table~\ref{tab:priors}. 
Since Ref.~\cite{LisHans} sets $L_4$ and $L_6$ exactly to zero, in accord with 
large $N_c$ expectations~\cite{L4largeNc}, for the width of $L_4$ and $2L_6-L_4$  
we use the error in the determination by the MILC collaboration in Ref.~\cite{LisMILC}. 
The exact choice of NLO LEC's priors has a minimal impact 
on our final result. We have checked several equally reasonable choices for 
treating these parameters including: using the errors in Ref.~\cite{LisHans} as 
prior widths instead of widths 2 times larger; using widths 10 times larger; 
fixing the LEC's (instead of leaving them as parameters of the fit) to the values 
in Ref.~\cite{LisHans}; fixing $L_{4,5,6}$ instead to the values determined 
in Ref.~\cite{LisMILC}, where different lattice light quantities were computed on 
the same ensembles as here; 
shifting the values of each individual $L_i$ within their corresponding error as 
determined in Ref.~\cite{LisHans}; and, keeping $L_4=0$ according to its large 
$N_c$ limit. The conclusion from these checks is that 
our data is unable to disentangle the individual values of the $L_i$'s, but 
the value of the form factor in the continuum limit and at the physical 
quark masses changes by less than 0.03\%  with any of the choices listed above.

\begin{center}
\begin{figure}[thb]
\includegraphics[width=0.6\textwidth,angle=-90]{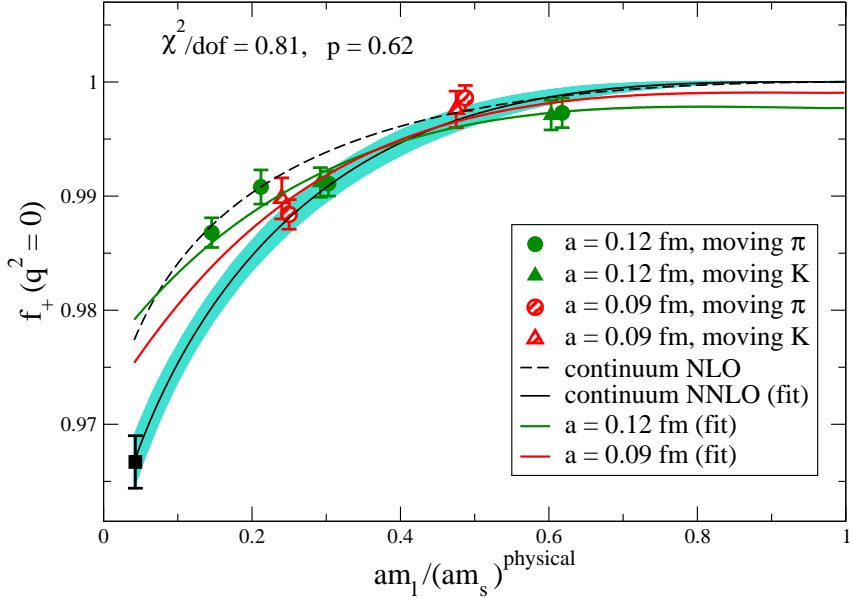}
\caption{Form factor $f_+(0)$ vs.~light-quark mass. Filled green (speckled red) symbols are the
$a\approx 0.12~{\rm fm}$ ($a\approx 0.09~{\rm fm}$) data points included in the chiral fit.
The dashed black line is the continuum NLO $\chi$PT prediction, $1+f_2$.
The solid black line is the extrapolation in the light-quark mass, keeping $m_s$
equal to its physical value, and turning off all discretization effects. The turquoise 
band shows the error for the continuum line. 
The green and red lines are also extrapolations in the
light-quark masses with $a\approx 0.12~{\rm fm}\,{\rm and}\,0.09~{\rm fm}$,
respectively. Errors in the data points are statistical only and were obtained
from 500 bootstrap samples. The differences between the green, red and solid black line 
reflect not only the discretization errors, but also the effect of the mistuned sea quark 
masses, which are particular large in the $a\approx0.12$ fm case. 
The error in the continuum value at the physical
quark masses includes also the uncertainty in the $\order(p^4)$ LEC's and
the HISQ hairpin parameters.
\label{fig:ChPTcentral}}
\end{figure}
\end{center}

The results from our preferred fitting strategy are shown in Fig.~\ref{fig:ChPTcentral}. 
The form factor in the continuum limit and at the physical quark masses is
\ba\label{eq:resultstat}
f_+(0) = 0.9667\pm0.0023\pm{\rm systematics}\,,
\ea
where the error, for now, is statistical plus the uncertainty 
from the errors of the hairpin parameters $\delta_{V,A}^{{\rm HISQ}}$ and the NLO LEC's. 
The p-value of the fit leading to the form factor in Eq.~(\ref{eq:resultstat}) and whose 
results are displayed in Fig.~\ref{fig:ChPTcentral} is 0.62. 
We note that approximately 15--25\% of the apparent discretization effects in the 
$0.12~{\rm fm}$ and $0.09~{\rm fm}$ extrapolations to the physical point 
(green and red lines in Fig.~\ref{fig:ChPTcentral}) are actually due to mistunings of 
the sea strange-quark mass.

\section{Systematic Error Analysis}

\label{sec:Errors}

In this section, we discuss the systematic errors in the calculation of $f_+(0)$ and the
different strategies followed to estimate them. The uncertainties discussed in this 
section are summarized in Table.~\ref{tab:errorbudget}.

\subsubsection{Chiral extrapolation and fit function}

\label{sec:fitfunction}

The fit function is completely determined at NLO in $\chi$PT, given the values of the 
taste-breaking parameters. In particular, the value of the decay constant that enters in 
$f_2^{PQ\chi PT}(a)$, the experimental value for $f_\pi$, is fixed by the two-loop formulae 
employed in our analysis. Any other choice for this parameter requires the use of a 
different NNLO expression. Hence, we can estimate the uncertainty associated with the choice 
of fit function and the truncation of the $\chi$PT series by exploring different choices 
we make for the NNLO description of the form factor.

First, we perform a set of chiral fits using the fit function in 
Eq.~(\ref{eq:ChPTfunction}), in which we describe NNLO and higher order 
chiral effects with an analytical parametrization. 
In order to compare with our result in 
Eq.~(\ref{eq:resultstat}) and disentangle discretization effects from the chiral 
extrapolation error, we parametrize the discretization effects in 
Eq.~(\ref{eq:ChPTfunction}) in the same way as for our preferred fit function 
using Eq.~(\ref{eq:ChPTtwoloop}): we take $K_2^{(a)}=0$ and $K_1^{(a)}\ne0$. 
We perform several fits in which we include the NNLO term $C^{(1)}_6$ along with 
one or two of the other terms in Eq.~(\ref{eq:ChPTfunction}). 

The fit function equivalent to the one we use in our preferred fit, namely,
the pure NNLO expression, which includes terms proportional to $K_1^{(a)}$ 
and $C_6^{(1)}$ in Eq.~(\ref{eq:ChPTfunction}), and choosing the physical value 
for the decay constant $f_\pi$, agrees with our central value in Eq.~(\ref{eq:resultstat}) 
within statistical errors. The difference between the determinations is $\sim0.15\%$, 
although using the two-loop $\chi$PT gives a better fit. If one includes the NNNLO term 
proportional to $(r_1m_\pi)^2$ in Eq.~(\ref{eq:ChPTfunction}), the quality of the 
fit improves and the result for $f_+(0)$ is nearly identical to Eq.~(\ref{eq:resultstat}). Including 
the logarithmic term proportional to $C_8^{(2)}$, instead of the term proportional 
to $C_8^{(1)}$, gives a value of $f_+(0)$ $\sim0.2\%$ lower than the one in 
Eq.~(\ref{eq:resultstat}) but with worse confidence level and much larger errors. 
Going beyond the inclusion of two chiral correction terms makes the fits unstable 
and considerably increases the statistical errors of the extrapolated point. 
With our data, we are not able to determine more than three coefficients, 
including the one proportional to the lattice spacing squared, $K_1^{(a)}$.

Another issue related to the use of the analytical parametrization in 
Eq.~(\ref{eq:ChPTfunction}) is the freedom to choose 
the value of the decay constant used at NLO, {\it i.e.}, in $f_2^{PQ\chi PT}$ (with 
the two-loop calculation, the freedom appears at NNLO). 
In our preferred NNLO $\chi$PT analysis, using $f_K$ instead of $f_\pi$ changes the  
central value by a negligible 0.02\%, while using $f_0=(113.6 \pm 3.6 \pm 
7.7)~{\rm MeV}$~\cite{fKoverfpi3} instead of $f_\pi$ increases 
the result only by 0.2\%. 

The last ambiguity in our fit function is related to the fact that 
we have strange quarks simulated with different masses 
in the valence and sea sectors. 
We are using the physical values for the valence strange-quark masses on
each ensemble, but the sea strange-quark masses have values that differ
from their physical values by a factor of $\sim 1.5$ on the  $a\approx0.12~{\rm fm}$ 
ensembles and $\sim1.2$ on the  $\approx0.09~{\rm fm}$ ensembles. We correct
for this mistuning using PQS$\chi$PT at one loop,
but we cannot correct at the two-loop level since the partially quenched
$\chi$PT expressions are not available. 
We obtain the value in Eq.~(\ref{eq:resultstat}) using the valence meson
masses in the two-loop $\chi$PT function $f_4$. 
If we use the sea meson masses instead, keeping the factor $(m_K^2-m_\pi^2)^2$ 
from the valence sector, the extrapolated result for $f_+(0)$ decreases by 0.3\%. 
At NLO, the difference between the correct $1+f_2^{PQS\chi PT}$ and the 
one taking all the masses equal to the valence masses is $0.2\%$ 
on the coarse ensembles and $0.07\%$ on the fine ensembles, and the 
NLO contribution is around 3 times larger than the NNLO. Thus, the 
0.3\% shift most likely overestimates the effect of the mistuning 
of $m_s$ in the sea. 

Finally, the addition of a NNNLO term of the form $(m_\pi^2-m_K^2)^2\times m_\pi^2$ 
to our fit function in Eq.~(\ref{eq:ChPTtwoloop})  results in a negligible shift in 
the extrapolated result for $f_+(0)$. 
In summary, all the effects described above provide estimates of the error associated with our 
choice of fit function. We take the largest shift in the central value, 0.3\%, 
as the systematic error.

\begin{table}[t]
    \centering
\caption{\label{tab:errorbudget} Complete error budget and total error for
$f_+(0)$. All errors are given in percent.}
\begin{tabular}{lc}
\hline
 \hline
Source of uncertainty &  Error $f_+(0)$ (\%) \\
\hline
Statistics & $0.24$ \\
Chiral extrapolation \& fitting & $0.3$ \\
Discretization & $0.1$ \\
Scale & $0.06$ \\
Finite volume & $0.1$  \\
\hline
Total Error & 0.42 \\
\hline\hline
\end{tabular}
\end{table}

\subsubsection{Discretization errors}

The dominant sources of discretization errors in our calculation are taste-violating 
and finite momentum effects. The leading order contribution from the 
first source is removed by using PQS$\chi$PT at one loop. The remaining taste-breaking 
errors are of $\order(a^2\alpha_s^3\Lambda_{QCD}^2,a^4\Lambda_{QCD}^4)$. 
The latter source yields errors of $\order(\alpha_s(ap)^2)$.

Since we have results only at two different values of the lattice spacing, we can 
reliably determine the value of one fit term that depends on the lattice spacing. 
For our central value, we choose the simplest parametrization of discretization effects, 
a term given by $(a/r_1)^2$ times a constant $K_1^{(a)}$---see Eq.~(\ref{eq:ChPTfunction}). 
We choose to allow this term to break the AG theorem since we expect errors of $\order((ap)^2)$ 
to be larger than $\order(a^2\alpha_s^2\Lambda_{QCD}^2,a^4\Lambda_{QCD}^4)$ errors. 
The value obtained from the fit for the $a^2$ 
coefficient, $K_1^{(a)}=-0.011(9)$, suggests that discretization errors after removing 
the dominant taste-breaking effects are very small.  
Indeed, our data are well described by a fit that includes only the one-loop PQS$\chi$PT 
terms and omits the $(a/r_1)^2$ term. With $K_1^{(a)} = 0$, we obtain a result that agrees with 
Eq.~(\ref{eq:resultstat}) at the 0.05\% level for both the central value and the error. 
However, we do need to include taste-breaking effects at one loop, since the statistical 
errors in our data are very small, 0.1--0.15\%.

We conclude that the remaining $\order(a^2)$ errors after the continuum extrapolation are 
negligible. Nevertheless, we try alternate parametrizations of the generic lattice-spacing 
dependence to further 
test the size of the remaining discretization effects. We perform fits in which we 
replace the $K_1^{(a)}$ term by (1) the $K_2^{(a)}$ term in Eq.~(\ref{eq:ChPTtwoloop}),   
(2) a term of order $(a/r_1)^2\,(m_K^2-m_\pi^2)$, (3) a term proportional to 
$\left(a/r_1\right)^2(r_1|\vec{p}_P|)^2$, where $P=\pi(K)$ for a moving pion (moving kaon), and 
(4) a term  proportional to $(a/r_1)^2(r_1q)^2$. The last of these tests is motivated by 
the observed deviations of the simulation $q^2$ from zero. 
These small non-zero values of the momentum transfer, $(r_1q)^2\sim 10^{-4}$, are due to momentum dependent  
discretization errors. None of the four fits increase the central value of Eq.~(\ref{eq:resultstat}) 
by more than 0.1\%. We therefore take 0.1\% as 
our estimate of the remaining discretization effects.

\subsubsection{Finite volume effects}

We carried out an additional simulation on an ensemble with the same parameters as the 
$a\approx 0.12~{\rm fm}$, $am_l=0.2am_s$ ensemble   
(second line in Table~\ref{tab:ensemblesa}) but with a larger volume 
(third line in Table~\ref{tab:ensemblesa}). 
This simulation in a larger volume gives a value for $f_+(0)$ which is 
$\sim 0.1\%$ lower. Note that the physical volumes of the other ensembles in 
Table~\ref{tab:ensemblesa}  are comparable to that of the $a\approx 0.12$~fm, 
$am_l = 0.2 am_s$ ensemble, and in particular the most chiral $a\approx 0.09$~fm 
ensemble has approximately equal size as the $0.12~{\rm fm}$ ensemble where the 
finite size effects were checked explicitly, and has roughly the same sea quark masses. 
We therefore take the 0.1\% difference as the estimate of the finite 
volume error.

\subsubsection{Lattice scale}

Since we are calculating a dimensionless quantity, $f_+(0)$, 
uncertainties from the conversion from lattice units to physical units 
are small. We perform this conversion through 
an intermediate scale, the $r_1$ scale~\cite{r11,r12}, by rewriting all dimensionful 
quantities entering in the chiral fit function in $r_1$ units. 
The lattice parameters are converted to $r_1$ units by 
using the values of $r_1/a$ listed in Table~\ref{tab:ensemblesa} for each MILC ensemble. 
The physical parameters are expressed in $r_1$ units using the determination of 
the physical $r_1$ in Ref.~\cite{r1HPQCD}, $r_1=(0.3133\pm0.0023)~{\rm fm}$. We use 
this value for consistency with the determination of the valence strange-quark masses 
in our simulations, which were tuned to their physical values in Ref.~\cite{r1HPQCD} 
using it. This value agrees within errors with the one obtained from 
the 2009 MILC determination of $r_1f_\pi$~\cite{fpiMILC09} combined with the PDG value 
of $f_\pi$~\cite{PDG2011}, $r_1=0.3117\pm0.0022$~\cite{decayconstants2011}. 

If we change the value of $r_1$ to the extremes allowed by its total error,  
the form factor $f_+(0)$ shifts 0.06\% from our central value. 
We add this shift as a systematic error associated with the choice of scale.

\subsubsection{Chiral scale}

We checked that our result is independent of the chiral scale when varying 
$\Lambda_\chi$ between $M_\rho\pm0.5~{\rm GeV}$. Indeed, we find it is independent 
up to small fluctuations 
due mainly to the fact that the sunset two-loop contribution is handled numerically 
by means of a lookup table. We thus do not need to add any extra uncertainty 
from the choice of scale.

\section{Calculation of the $\order(p^6)$ LEC combination $C_{12}^r+C_{34}^r$}

\label{sec:p6LECs}

\begin{table}[t]
    \centering
\caption{\label{tab:errorbudgetcij} Complete error budget and total error for
the combination of $\order(p^6)$ LEC's $C_{12}^r+C_{34}^r$.
All errors are given in percent.}
\begin{tabular}{lcc}
\hline
 \hline
Source of uncertainty & $(C_{12}^r+C_{34}^r)$ (\%) \\
\hline
Statistics & $9.6$ \\
Chiral extrapolation \& fitting & $12$ \\
Discretization & $5.7$ \\
Scale & $0.2$ \\
Finite volume & $\sim 11$ \\
\hline
Total Error &  19.7 \\
\hline\hline
\end{tabular}
\end{table}

As a byproduct of our calculation, from the parameter $C_6'^{(1)}$ in 
Eq.~(\ref{eq:ChPTtwoloop}) we can extract the combination of $\order(p^4)$ and 
$\order(p^6)$ LEC's
\ba\label{eq:cijsplusl5}
(C_{12}^r+C_{34}^r-(L_5^r)^2)(M_\rho) = (3.62\pm0.33)\times10^{-6}\,,
\ea
where the error is statistical only. Using the result in Eq.~(\ref{eq:cijsplusl5}) 
and the LEC $L_5^r(M_\rho)=(0.97\pm0.22)\times10^{-3}$, which we also 
get as an output of our chiral fit (although it is the same as the corresponding
prior within the given precision), we obtain 
\ba\label{eq:cijs}
(C_{12}^r+C_{34}^r)(M_\rho) = (4.57\pm0.44\pm0.90)\times10^{-6}\,,
\ea
where the first error is statistical and also includes the uncertainty 
from the errors of the hairpin parameters $\delta_{V,A}^{HISQ}$ and the NLO LEC's, and 
the second error is from the systematic errors added in quadrature. 
The detailed error budget for this quantity is given in Table~\ref{tab:errorbudgetcij}. 
In this case, we do not vary the value of the decay constant used in the chiral
expansion at NNLO to estimate the systematic error associated with the chiral
extrapolation and choice of fit function, since the LEC's are not physical quantities
and depend on convention. The value in Eq.~(\ref{eq:cijs}) should be used taking into
account the type of fit performed for its extraction. We obtain the other systematic
errors in the same way as for $f_+(0)$, with the exception of finite volume effects.
We estimate that error to be of the same size as that for $1-f_+(0)$, which
highly enhances the effect as compared to $f_+(0)$. 

The result in Eq.~(\ref{eq:cijs}) agrees with non-lattice determinations in 
Refs.~\cite{Ciriglianof+,Jaminf+,LR84}. The authors of those papers used the 
large $N_c$ approximation, a coupled-channel dispersion relation analysis that 
provides the slope and the curvature of the scalar $K\pi$ form factor, and a quark model, 
respectively,  to calculate the contribution from the combination $C_{12}+C_{34}$ to the form 
factor $f_+(0)$. The calculation of $C_{12}+C_{34}-L_5^2$ in Ref.~\cite{Kastner:2008ch}, 
based on $\chi$PT, large $N_c$ estimates of the LEC's, and dispersive methods 
is $\sim3\sigma$ smaller than our result in Eq.~(\ref{eq:cijsplusl5}).

\section{Discussion of results and future improvements}

\label{sec:Results}

We summarize the error budget and total error 
for our calculation of $f_+(0)$ discussed in Sec.~\ref{sec:Errors} 
in Table~\ref{tab:errorbudget}. Our final result is 
\ba\label{eq:result}
f_+(0) = 0.9667\pm 0.0023\pm0.0033=0.9667\pm0.0040\, ,
\ea
where the first error in the middle expression is statistical and the second is the sum in 
quadrature of the different systematic errors. 
In Table~\ref{tab:fpcomparison}, we list our result together with other determinations 
from both unquenched lattice methods and different analytical approaches.  
Our result is about 1$\sigma$ larger than previous unquenched 
lattice determinations by the RBC/UKQCD and ETMC Collaborations, and 
has somewhat smaller errors mainly due to the use of two-loop $\chi$PT 
in the chiral extrapolation and the use of data at two lattice spacings. 
On the other hand, the form factor in Eq.~(\ref{eq:result}) is smaller than 
those coming from analytical approaches based on the use of 
two-loop $\chi$PT and model estimates of the $\order(p^6)$ LEC's, although 
compatible within errors (except for the calculation in Ref.~\cite{Kastner:2008ch}). 

\begin{table}[t]
\caption{Form factor $f_+(0)$ as extracted from unquenched lattice calculations (first half
of the table), phenomenological approaches based on the use of two-loop $\chi$PT, and the pioneering
calculation by Leutwyler and Roos using a quark model. For those calculations based on 
two-loop $\chi$PT, we also indicate the method used in the estimate of the $\order(p^6)$ LEC's.
\label{tab:fpcomparison} }
\begin{tabular}{cccc}
\hline\hline
Group & $f_+(0)$ & Method & Sea content\\
\hline
This work & $0.9667(23)(33)$ & staggered fermions & $N_f=2+1$ \\
RBC/UKQCD~\cite{Boyle:2010bh} & $0.9599(34)\left({}^{+31}_{-43}\right)$ & DW fermions & $N_f=2+1$ \\
ETMC~\cite{f+ETMC}     & $0.9560(57)(62)$ & twisted mass fermions & $N_f=2$\\
\hline
Kastner \& Neufeld~\cite{Kastner:2008ch} & $0.986(8)$ &  $\chi$PT + large $N_c$ limit + dispers. & -- \\
Cirigliano {\it et al.}~\cite{Ciriglianof+} & $0.984(12)$ & $\chi$PT + large $N_c$ limit & -- \\
Jamin, Oller, \& Pich~\cite{Jaminf+} & $0.974(11)$ & $\chi$PT + disp.(scalar form factor)  & -- \\
Bijnens \& Talavera~\cite{BT03} & $0.976(10)$ & $\chi$PT + Leutwyler \& Roos &  -- \\
\hline
Leutwyler \& Roos~\cite{LR84} & $0.961(8)$ & Quark model & -- \\
\hline\hline
\end{tabular}
\end{table}                                                                               

\begin{center}
\begin{figure}[tb]
\includegraphics[angle=-90,width=0.8\textwidth]{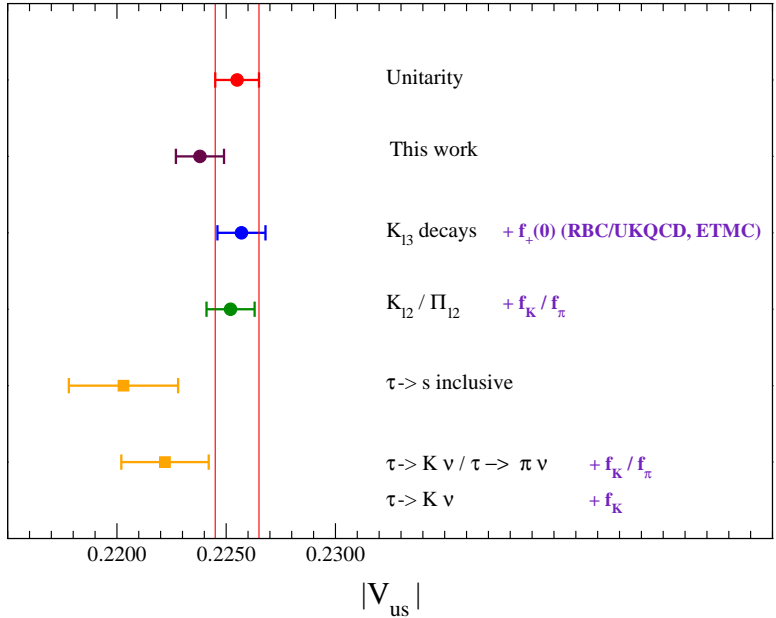}
\caption{Comparison of $\vert V_{us}\vert$ as extracted from $K$ leptonic, 
$K$ semileptonic, $\tau$ decays, and the value obtained in this work. Theoretical 
inputs for $f_+(0)$, $f_K/f_\pi$ and $f_K$ are taken from Refs.~\cite{Boyle:2010bh,f+ETMC},  
\cite{lataver,fKoverfpi1,fKoverfpi2,fKoverfpi3,fKoverfpi4,fKoverfpi5}, and 
\cite{lataver,fKoverfpi1,fKoverfpi3,fK1,fK2}, respectively. The unitarity value of 
$\vert V_{us}\vert$ is obtained from $\vert V_{ud}\vert=0.97425(22)$~\cite{Vud08} and 
neglecting the contribution from $\vert V_{ub}\vert$. The vertical lines correspond to 
the unitarity prediction. 
See text for further explanation.\label{fig:vus}}
\end{figure}
\end{center}

With the form factor in Eq.~(\ref{eq:result}) and the latest average of 
experimental measurements of $K$ semileptonic decays~\cite{MoulsonCKM12}, 
$\vert V_{us}\vert f_+(0) = 0.2163(5)$, 
we obtain the following value for the CKM matrix element $\vert V_{us}\vert$: 
\ba\label{eq:Vus}
\vert V_{us}\vert = 0.2238\pm0.0009\pm 0.0005 = 0.2238\pm0.0011\,, 
\ea
where the first error is from $f_+(0)$ and the second one is 
experimental. Using the average over superallowed nuclear beta decay determinations, 
$\vert V_{ud}\vert=0.97425(22)$~\cite{Vud08},  and neglecting the value of 
$\vert V_{ub}\vert$, the result in Eq.~(\ref{eq:Vus}) allows us to check
unitarity in the first row of the CKM matrix, 
\ba
\Delta_{{\rm CKM}}\equiv \vert V_{ud}\vert^2 + \vert V_{us}\vert^2 + \vert V_{ub}\vert^2 - 1 =
-0.0008(6)\,.
\ea
In Fig.~\ref{fig:vus} we summarize the status of the determination of $\vert V_{us}\vert$ 
from different sources: this work, semileptonic decays using as theory input previous unquenched 
lattice calculations of $f_+(0)$ listed in Table~\ref{tab:fpcomparison}, leptonic 
decays using as input the average over $N_f=2+1$ lattice calculations of 
$f_K/f_\pi$~\cite{lataver,fKoverfpi1,fKoverfpi2,fKoverfpi3,fKoverfpi4,fKoverfpi5}, inclusive hadronic 
$\tau$ decays~\cite{tauHFAG,tauVus},\footnote{The value shown in Fig.~\ref{fig:vus} 
uses the theoretical calculation in \cite{tauVus} together with the latest experimental 
averages for hadronic $\tau$ decay modes~\cite{tauHFAG}, except for a few modes for which the 
experimental measurements are substituted by theoretical predictions as described in 
Ref.~\cite{Passemartau2012}.} and exclusive $\tau$ decay modes~\cite{tauHFAG} 
that require either 
$f_K/f_\pi$~\cite{lataver,fKoverfpi1,fKoverfpi2,fKoverfpi3,fKoverfpi4,fKoverfpi5} or 
$f_K$~\cite{lataver,fKoverfpi1,fKoverfpi3,fK1,fK2} from the lattice 
as theory inputs. The value predicted by unitarity of the CKM matrix is also included 
in the figure. 

Using staggered fermions we have obtained the most precise lattice-QCD value of 
$f_+(0)$ to date. Three elements of our work are the key to this precision.
First, the MILC asqtad ensembles yield very small statistical errors,  
$\sim 0.1-0.15\%$, providing a solid foundation for the whole analysis. 
Second, using the HISQ action for the valence quarks leads to very small discretization 
errors. Third, the $\chi$PT description of the chiral behavior of our data---including 
taste-breaking, partially quenched, and mixed-action effects at one loop---permit a 
controlled extrapolation to the continuum limit and physical light-quark mass.
In particular, our analysis shows that one-loop PQS$\chi$PT may be enough to account for 
the discretization effects observed in our data, so our results do not change even if 
we disregard any additional sources of discretization error. 

The uncertainty in the form factor still dominates the determination of 
$\vert V_{us}\vert$ from $K$ semileptonic decays, so further improvements 
are necessary. We are in the process of reducing the dominant sources of systematic error in our analysis 
by performing simulations with physical light- and strange-quark masses on the HISQ 
$N_f=2+1+1$ MILC ensembles~\cite{hisqconfig}. With lattice data at the physical quark 
masses we expect to greatly reduce the chiral extrapolation errors. 
Discretization errors are also 
considerably smaller for the HISQ action than for the asqtad action, as can be seen 
explicitly in Ref.~\cite{latt2012}. Another advantage of the HISQ ensembles is that 
the strange sea-quark masses are much better tuned than in the asqtad ones, which will 
reduce the dominant contribution to the chiral-fit systematic error. Finally, the MILC 
HISQ ensembles include the effects of charm quarks in the sea.

\begin{acknowledgments}

We thank Jon Bailey, Johan Bijnens, Christine Davies, Eduardo Follana, Pere Masjuan, 
and Heechang Na  for useful discussions. 
We thank Johan Bijnens for making his NLO partially quenched $\chi$PT and NNLO
full QCD $\chi$PT codes available to us. We thank Jon Bailey for the careful reading 
of this manuscript. 
Computations for this work were carried out with resources provided by
the USQCD Collaboration, the Argonne Leadership Computing Facility,
the National Energy Research Scientific Computing Center, and the Los Alamos 
National Laboratory, which are funded by the Office of Science of the United 
States Department of Energy; and with resources provided by the National Institute 
for Computational Science, the Pittsburgh Supercomputer Center, 
the San Diego Supercomputer Center, and the Texas Advanced Computing Center,
which are funded through the National Science Foundation's Teragrid/XSEDE Program.
This work was supported in part by the U.S. Department of Energy under Grants
No.~DE-FG02-91ER40628 (C.B.),
No.~DOE~FG02-91ER40664 (Y.M.),
No.~DE-FC02-06ER41446 (C.D., J.F., L.L., M.B.O.),
No.~DE-FG02-91ER40661 (S.G., R.Z.),
No.~DE-FG02-91ER40677 (D.D., A.X.K.),
No.~DE-FG02-04ER-41298 (J.K., D.T.);
by the National Science Foundation under Grants
No.~PHY-1067881, No.~PHY-0757333, No.~PHY-0703296 (C.D., J.F., L.L., M.B.O.),
No.~PHY-0757035 (R.S.); 
by the Science and Technology Facilities Council and the Scottish Universities 
Physics Alliance (J.L.);
by the MICINN (Spain) under grant FPA2010-16696 and Ram\'on y Cajal program (E.G.);
by the Junta de Andaluc\'ia (Spain) under Grants No.~FQM-101, No.~FQM-330, and 
No.~FQM-6552 (E.G.); 
and by European Commission (EC) under Grant No.~PCIG10-GA-2011-303781 (E.G.). 
This manuscript has been co-authored by employees of Brookhaven Science Associates, 
LLC, under Contract No.~DE-AC02-98CH10886 with the U.S. Department of Energy.
Fermilab is operated by Fermi Research Alliance, LLC, under Contract 
No.~DE-AC02-07CH11359 with the U.S. Department of Energy.

\end{acknowledgments}

\appendix

\section{Staggered Chiral Perturbation Theory at NLO}

\label{app:SCHPT}

As described in Sec.~\ref{sec:chiralfits}, we use S$\chi$PT 
to incorporate the dominant discretization effects at one loop.  
For staggered quarks the taste-nonsinglet pseudoscalar mesons masses are split 
according to five taste representations $\Xi=P,A,T,V,I$: 
\ba\label{eq:mesonmass}
M_{ij,\Xi}^2 = \mu(m_i+m_j) + a^2\Delta_\Xi\,,
\ea
with $m_i$ and $m_j$ the quark masses. All the taste-nonsinglet 
pseudoscalar mesons contribute via the loops in the S$\chi$PT calculation, even  
when we choose our external states to be the ones with $\Xi=P$. 

Since we are using a different staggered action in the 
sea (asqtad) and in the valence (HISQ) sectors, there are mixed-action effects that 
we need to incorporate in our expressions. At one loop, those effects arise as 
changes to the taste splittings $\Delta_{\Xi}$, extra 
terms in the disconnected propagators, and the appearance of three different hairpin 
coefficients in each channel. These modifications affect the structure of the 
disconnected propagator only in the axial and vector channels, changing them from the 
usual form~\cite{SCHPT}
\ba 
{\cal D}^{\Xi}_{XY}(p) = -\frac{a^2\delta_{\Xi}}{(p^2+m_{X_\Xi}^2)(p^2+m_{Y_\Xi}^2)}
\frac{(p^2+m_{U_\Xi}^2)(p^2+m_{D_\Xi}^2)(p^2+m_{S_\Xi}^2)}
{(p^2+m_{\pi_\Xi}^2)(p^2+m_{\eta_\Xi}^2)(p^2+m_{\eta'_\Xi}^2)}\, ,
\ea
with $\Xi=V,A$ and $U,D,S$ the $u\bar u$, $d\bar d$, and $s\bar s$ sea-quark mesons 
respectively, to the following form~\cite{KtopilnuSChPT}
\ba
{\cal D}^{\Xi}_{XY}(p) & = & -\frac{a^2(\delta^{v\sigma}_{\Xi})^2/
\delta^{\sigma\sigma}_\Xi}{(p^2+m_{X_\Xi}^2)(p^2+m_{Y_\Xi}^2)}
\frac{(p^2+m_{U_\Xi}^2)(p^2+m_{D_\Xi}^2)(p^2+m_{S_\Xi}^2)}
{(p^2+m_{\pi_\Xi}^2)(p^2+m_{\eta_\Xi}^2)(p^2+m_{\eta'_\Xi}^2)}\nonumber\\
&& -\frac{a^2[\delta^{vv}-(\delta^{v\sigma}_{\Xi})^2/\delta^{\sigma\sigma}_\Xi]}
{(p^2+m_{X_\Xi}^2)(p^2+m_{Y_\Xi}^2)}\,
\ea
where $v$ and $\sigma$ label the valence and sea quark, respectively.  
It seems reasonable to expect $\delta_\Xi^{v\sigma} \sim \sqrt{\delta_\Xi^{vv}
\delta_\Xi^{\sigma\sigma}}$, so the mixed action effects should be small. 
To take this into account in our fits we define the parameter $\delta_\Xi^{{\rm mix}} =  
\delta^{vv}-(\delta^{v\sigma}_{\Xi})^2/\delta^{\sigma\sigma}_\Xi$ and write the 
S$\chi$PT expressions in terms of $\delta_\Xi^{{\rm mix}}$ (that we expect to be 
suppressed with respect to the parameters in the sea and the valence sectors) 
and $\delta^{vv}_\Xi$:
\begin{eqnarray}
{\cal D}^{\Xi}_{XY}(p) & = & -\frac{a^2(\delta^{vv}_\Xi-\delta^{{\rm mix}}_\Xi)}
{(p^2+m_{X_\Xi}^2)(p^2+m_{Y_\Xi}^2)}
\frac{(p^2+m_{U_\Xi}^2)(p^2+m_{D_\Xi}^2)(p^2+m_{S_\Xi}^2)}
{(p^2+m_{\pi_\Xi}^2)(p^2+m_{\eta_\Xi}^2)(p^2+m_{\eta'_\Xi}^2)}\nonumber\\
&& -\frac{a^2\delta^{{\rm mix}}_\Xi}{(p^2+m_{X_I}^2)(p^2+m_{Y_I}^2)}\,\,.
\end{eqnarray}
Although we can not determine very precisely the value of $\delta^{{\rm mix}}_\Xi$ from 
our data, the chiral fits prefer non-zero values that are the same sign as 
but an order of magnitude smaller than $\delta^{vv}_\Xi$. After these 
mixed-action expressions were derived, we discovered that they had been previously 
worked out in Ref.~\cite{Bae:2010ki}, including the comments about the expected size 
of the $\delta^{v\sigma}_\Xi$.

The partially quenched one-loop calculation involves mesons made of two valence, two 
sea, and one valence and one sea quarks. The corresponding masses are given by the 
definition in (\ref{eq:mesonmass}) with the taste splittings $\Delta_\Xi$ being 
$\Delta_\Xi^{HISQ}$, $\Delta_\Xi^{asqtad}$, and $\Delta_\Xi^{{\rm mix}}$, respectively.

\subsection{Result for $f_2$}

A detailed description of the one-loop PQS$\chi$PT calculation with and without a mixed 
action setup  will be given elsewhere~\cite{KtopilnuSChPT}. Here, we just give the final 
result needed for the extrapolation of the form factor $f_+(q^2=0)$. 
The NLO PQS$\chi$PT expression for the form factor at zero momentum transfer 
in the isospin limit and for $N_f=2+1$ flavors of sea quarks, 
$f_2(0)\equiv f_2^{PQ\chi PT}(q^2=0,a)$, is (for $x$ and $y$ the light and strange 
valence quarks respectively)

\begin{eqnarray}\label{eq:2p1_f2tot}
  f_2(0) & =&
\frac{-1}{2(4\pi f)^2}\Biggl\{
  \frac{1}{16}\sum_{f,\Xi}
  \left[
  \ell\left(m_{xf,\Xi}\right)
  +\ell\left(m_{yf,\Xi}\right)
  \right]
  +\frac{1}{3}\Biggl[\sum_{j}
    \frac{\partial}{\partial m_{X,I}^2}
  \left(
  R^{[2,2]}_{j} \left(\cM^{(2,x)}_I ; \mu^{(2)}_I\right)
  \ell(m_{j,I})\right)\nonumber \\* &&
  +\sum_{j}
    \frac{\partial}{\partial m_{Y,I}^2}
    \left(
    R^{[2,2]}_{j}  \left(\cM^{(2,y)}_I ; \mu^{(2)}_I\right)
    \ell(m_{j,I})\right) +2\sum_{j}
  R^{[3,2]}_{j} \left(\cM^{(3,xy)}_I ; \mu^{(2)}_I\right)
  \ell(m_{j,I})\Biggr]
  \nonumber \\* 
&&+\frac{1}{4}\sum_{f,\Xi} \, 
B_{22}(m_{xf,\Xi},m_{yf,\Xi},0)
+ \frac{4}{3}\;\frac{\partial}{\partial m_{X,I}^2} \left\lbrace 
\sum_j \tilde B_{22}(m_{xy,I}^2,m_{j,I}^2,0) 
R^{[2,2]}_{j} 
\left(\cMI^{(2,x)} ; \mu^{(2)}_I\right)\right\rbrace\
\nonumber\\ 
&& + \frac{4}{3}\;\frac{\partial}{\partial m_{Y,I}^2} \left\lbrace 
\sum_j \tilde B_{22}(m_{xy,I}^2,m_{j,I}^2,0) 
R^{[2,2]}_{j} 
\left(\cMI^{(2,y)} ; \mu^{(2)}_I\right)\right\rbrace\
\nonumber\\ 
&&+\frac{8}{3}\; \sum_{j}\, 
\tilde B_{22}(m_{xy,I}^2,m_{j,I}^2,0) R^{[3,2]}_j \left(\cMI^{(3,xy)} ; 
\mu^{(2)}_I\right) \nonumber\\
&&+ a^2 (\delta_V^{vv}-\delta_V^{{\rm mix}})
\Biggl[
    \sum_{j}
    \frac{\partial}{\partial m_{X,V}^2}
  \left(
  R^{[3,2]}_{j}  \left(\cM^{(3,x)}_V ; \mu^{(2)}_V\right)
  \ell(m_{j,V})
  \right)\nonumber \\* &&
  +\sum_{j}
    \frac{\partial}{\partial m_{Y,V}^2}
    \left(
    R^{[3,2]}_{j}\left(\cM^{(3,y)}_V ; \mu^{(2)}_V\right)
    \ell(m_{j,V})\right)\nonumber
  +2\sum_{j}
  R^{[4,2]}_{j}\left(\cM^{(4,xy)}_V ; \mu^{(2)}_V\right)
  \ell(m_{j,V})
    \nonumber \\* 
&&  
+ 4\; \frac{\partial}{\partial m_{X,V}^2}
\left\lbrace 
\sum_j \tilde B_{22}(m_{xy,V}^2,m_{j,V}^2,0) 
R^{[3,2]}_{j} \left(\cMV^{(3,x)} 
; \mu^{(2)}_V\right) \right\rbrace
\\ 
&& +4\;\frac{\partial}{\partial m_{Y,V}^2}
\left\lbrace 
\sum_j \tilde B_{22}(m_{xy,V}^2,m_{j,V}^2,0) 
R^{[3,2]}_{j} \left(\cMV^{(3,y)} 
; \mu^{(2)}_V\right) \right\rbrace
 \nonumber\\ 
&&+8\; \sum_{j}\, 
\tilde B_{22}(m_{xy,V}^2,m_{j,V}^2,0) R^{[4,2]}_j \left(\cMV^{(4,xy)} ; 
\mu^{(2)}_V\right) \,\Biggl\rbrack\nonumber\\
&&+ a^2\delta_V^{{\rm mix}}
\Biggl\lbrack\frac{\partial \ell(m_{X,V}) }{\partial m_{X,V}^2}
+ \frac{\partial \ell(m_{Y,V}) }{\partial m_{Y,V}^2} 
+2\; \frac{\left(\ell(m_{X,V}) -\ell(m_{Y,V})\right)}{m_{Y,V}^2-m_{X,V}^2}
\nonumber\\*
&&+  4\; \frac{\partial}{\partial m_{X,V}^2}\tilde B_{22}(m_{xy,V}^2,m_{X,V}^2,0)
+4\; \frac{\partial}{\partial m_{Y,V}^2}\tilde B_{22}(m_{xy,V}^2,m_{Y,V}^2,0)
\nonumber\\
&&+\;\frac{8}{m_{Y,V}^2-m_{X,V}^2}\left(\tilde B_{22}(m_{xy,V}^2,m_{X,V}^2,0)
-\tilde B_{22}(m_{xy,V}^2,m_{Y,V}^2,0)\right)
\Biggr\rbrack\nonumber\\
&& +  \Bigl[ V \to A \Bigr]  \Biggl\}\ , \nonumber 
\end{eqnarray}
where $\Xi$ runs over the sixteen independent meson tastes, $f$ runs
over sea-quark flavors, $\ell(m)$ and $\tilde B_{22}$ are chiral logarithm functions 
defined below, and $R_j^{[n,k]}(\cM;\mu)$ are residue functions introduced in 
Ref.~\cite{SCHPT}, with
$\cM$ and $\mu$ various sets of meson masses:
\begin{eqnarray}\label{eq:2p1_denom_mass_sets}
        \{\cM^{(2,z)}_\Xi\}& \equiv & \{m_{\eta,\Xi}\,,
                m_{Z,\Xi}   \}\ , \nonumber \\*
        \{\cM^{(3,z,z')}_\Xi\}& \equiv & \{m_{\eta,\Xi},\,
        m_{Z,\Xi},\,m_{Z',\Xi}\}\ , \nonumber \\*
        \{\cM^{(3,z)}_\Xi\}& \equiv & \{m_{\eta,\Xi},\,m_{\eta',\Xi},\,
                m_{Z,\Xi}   \}\ , \nonumber \\*
        \{\cM^{(4,z,z')}_\Xi\}& \equiv & \{m_{\eta,\Xi},\,m_{\eta',\Xi},\,
        m_{Z,\Xi},\,m_{Z',\Xi}\}\ , \\*
        \{\mu^{(2)}_\Xi\}& \equiv & \{m_{U,\Xi},m_{S,\Xi}
                \}\ . \nonumber 
\end{eqnarray}
Here $z$ can be either $x$ or $y$, and $Z$ is the corresponding
$z\bar z$ meson ($X$ or $Y$).  

The chiral logarithm function $\ell$ is given by
\begin{equation}\label{eq:ell}
\ell(m)\equiv m^2\ln(m^2/\Lambda_\chi^2)\ ,
\end{equation} 
with $\Lambda_\chi$ the chiral scale. The function
$\tilde B_{22}(m_1^2, m_2^2, q^2)$ is related to the function $\bar B_{22}$
defined in Ref.~\cite{Bijnens:2002hp} by
\begin{equation}\label{eq:B22}
\tilde B_{22}(m_1^2, m_2^2, q^2) = (4\pi)^2\;\bar B_{22}(m_1^2, m_2^2, q^2)\ .
\end{equation}
In the special case of $q^2=0$, $\tilde B_{22}$ takes the simple form
\begin{equation}\label{eq:B22q0}
\tilde B_{22}(m_1^2, m_2^2, 0) = -\frac{1}{4}\left(\frac{
m_2^2\,\ell(m_2^2) - m_1^2\,\ell(m_1^2)}{m_2^2-m_1^2}\right)+\frac{1}{8}\left(m_1^2
+ m_2^2\right)\ .
\end{equation}

It is not difficult to check 
that Eq.~(\ref{eq:2p1_f2tot}) obeys the Ademollo-Gato theorem~\cite{AGtheorem} 
in the valence masses and vanishes as $(m_y-m_x)^2$ as $m_y\to m_x$.
To show this one needs the identities obeyed by the $R_j^{[n,k]}$ \cite{SCHPT}
(or more simply, the structure of the integrals from which they arise), as well
as Eqs.~(\ref{eq:ell}) and (\ref{eq:B22q0}) for the terms involving the
sum over $f$. Eq.~(\ref{eq:2p1_f2tot}) analytically 
agrees with the isospin symmetric partially quenched continuum 
calculation~\cite{Bijnenspc}.


\begin{thebibliography}{150}

\bibitem{Cirigliano10}
 V.~Cirigliano, J.~Jenkins and M.~Gonz\'alez-Alonso,
  Nucl.\ Phys.\ B {\bf 830} (2010) 95
  [arXiv:0908.1754 [hep-ph]].

\bibitem{Cirigliano11}
 V.~Cirigliano, G.~Ecker, H.~Neufeld, A.~Pich and J.~Portol\'es,
  Rev.\ Mod.\ Phys.\  {\bf 84}, 399 (2012)
  [arXiv:1107.6001 [hep-ph]].

\bibitem{tauHFAG}
 Y.~Amhis {\it et al.}  [Heavy Flavor Averaging Group],
  arXiv:1207.1158 [hep-ex].


\bibitem{fKoverfpi1}
J.~Laiho and R.~S.~Van de Water,
  PoS LATTICE {\bf 2011}, 293 (2011)
  [arXiv:1112.4861 [hep-lat]]; 

\bibitem{fKoverfpi2}
S.~D\"urr {\it et al.}, 
  Phys.\ Rev.\ D {\bf 81}, 054507 (2010)
  [arXiv:1001.4692 [hep-lat]]; 

\bibitem{Aubin:2004fs}
  C.~Aubin {\it et al.}  [MILC Collaboration],
  Phys.\ Rev.\ D {\bf 70}, 114501 (2004)
  [hep-lat/0407028].

\bibitem{fKoverfpi3}
A.~Bazavov {\it et al.}  [MILC Collaboration],
  PoS LATTICE {\bf 2010}, 074 (2010)
  [arXiv:1012.0868 [hep-lat]]; 

\bibitem{fKoverfpi4}
 Y.~Aoki {\it et al.}  [RBC and UKQCD Collaborations],
  Phys.\ Rev.\ D {\bf 83}, 074508 (2011)
  [arXiv:1011.0892 [hep-lat]]; 

\bibitem{fKoverfpi5}
 E.~Follana {\it et al.}  [HPQCD and UKQCD Collaborations],
  Phys.\ Rev.\ Lett.\  {\bf 100}, 062002 (2008)
  [arXiv:0706.1726 [hep-lat]].

\bibitem{fKoverfpi6}
R.~Arthur {\it et al.}  [RBC and UKQCD Collaborations],
  arXiv:1208.4412 [hep-lat].

\bibitem{Bazavov:2013cp} 
A.~Bazavov {\it et al.} [MILC Collaboration], 
  arXiv:1301.5855 [hep-ph].

\bibitem{FLAG}
 G.~Colangelo {\it et al.},
  Eur.\ Phys.\ J.\ C {\bf 71}, 1695 (2011)
  [arXiv:1011.4408 [hep-lat]].

\bibitem{lataver}
J.~Laiho, E.~Lunghi and R.~S.~Van de Water,
  Phys.\ Rev.\ D {\bf 81} (2010) 034503
  [arXiv:0910.2928 [hep-ph]]. Updated information can be found in http://www.latticeaverages.org.

\bibitem{Marciano04}
W.~J.~Marciano,
  Phys.\ Rev.\ Lett.\  {\bf 93}, 231803 (2004)
  [hep-ph/0402299].

\bibitem{Antonellietal2010}
M.~Antonelli {\it et al.},
  Eur.\ Phys.\ J.\  {\bf C69}, 399-424 (2010).
  [arXiv:1005.2323 [hep-ph]].

\bibitem{MoulsonCKM12}
 M.~Moulson,
  arXiv:1209.3426 [hep-ex].
                                                 
\bibitem{Vud08}
J.~C.~Hardy and I.~S.~Towner,
  Phys.\ Rev.\ C {\bf 79}, 055502 (2009)
  [arXiv:0812.1202 [nucl-ex]].

\bibitem{Hisqaction}
 E.~Follana {\it et al.}  [HPQCD Collaboration],
  Nucl.\ Phys.\ Proc.\ Suppl.\  {\bf 129}, 447 (2004)
  [hep-lat/0311004];
 E.~Follana {\it et al.}  [HPQCD and UKQCD Collaborations],
 Nucl.\ Phys.\ Proc.\ Suppl.\  {\bf 129\&130}, 384 (2004) 
  [hep-lat/0406021];
 E.~Follana {\it et al.}  [HPQCD and UKQCD Collaborations],
  Phys.\ Rev.\ D {\bf 75}, 054502 (2007)
  [hep-lat/0610092].

\bibitem{lepage-1999-59}
  G.~P.~Lepage,
  Phys.\ Rev.\  D {\bf 59}, 074502 (1999)
  [arXiv:hep-lat/9809157].

\bibitem{Boyle:2010bh}
 P.~A.~Boyle {\it et al.} [RBC-UKQCD Collaboration],
 Eur.\ Phys.\ J.\  {\bf C69 } (2010)  159-167.
 [arXiv:1004.0886 [hep-lat]].

\bibitem{f+ETMC}
 V.~Lubicz {\it et al.} [ETM Collaboration],
  Phys.\ Rev.\  {\bf D80 } (2009)  111502.
  [arXiv:0906.4728 [hep-lat]];

\bibitem{HPQCD_Dtopi}
H.~Na, C.~T.~H.~Davies, E.~Follana, G.~P.~Lepage, J.~Shigemitsu [HPQCD Collaboration],
  Phys.\ Rev.\  {\bf D82 } (2010)  114506.
  [arXiv:1008.4562 [hep-lat]].


\bibitem{latt2010}
 J.~A.~Bailey {\it et al.} [Fermilab Lattice and MILC Collaboration],
  PoS {\bf LATTICE2010 } (2010)  306.
  [arXiv:1011.2423 [hep-lat]].

\bibitem{latt2011}
 E.~G\'amiz, C.~DeTar, A.~X.~El-Khadra, A.~S.~Kronfeld, P.~B.~Mackenzie and J.~Simone 
[Fermilab Lattice and MILC Collaboration],
  PoS LATTICE {\bf 2011}, 281 (2011)
  [arXiv:1111.2021 [hep-lat]].
                                                              
\bibitem{latt2012}
 E.~G\'amiz {\it et al.} [Fermilab Lattice and MILC Collaboration],
PoS LATTICE {\bf 2012}, 113 (2012)
  [arXiv:1211.0751 [hep-lat]].

\bibitem{Kanekolat12}
 T.~Kaneko  {\it et al.}  [JLQCD Collaboration],
  PoS LATTICE {\bf 2012}, 111 (2012)
  [arXiv:1211.6180 [hep-lat]].

\bibitem{Sivalingamlat12}
 P.~A.~Boyle, J.~M.~Flynn, A.~J\"uttner, C.~Sachrajda, K.~Sivalingam and J.~M.~Zanotti,
  arXiv:1212.3188 [hep-lat].

\bibitem{HPQCD09}
H.~Na, C.~T.~H.~Davies, E.~Follana, P.~Lepage and J.~Shigemitsu,
  PoS LAT {\bf 2009} (2009) 247
  [arXiv:0910.3919 [hep-lat]].

\bibitem{Koponen:2012di}
J.~Koponen {\it et al.}  [HPQCD Collaboration],
  arXiv:1208.6242 [hep-lat].

\bibitem{McNeile:2006bz}
  C.~McNeile {\it et al.}  [UKQCD Collaboration],
  Phys.\ Rev.\ D {\bf 73}, 074506 (2006)
  [hep-lat/0603007].
\bibitem{Bedaque:2004ax}
 P.~F.~Bedaque, J.~-W.~Chen,
 Phys.\ Lett.\  {\bf B616}, 208-214 (2005).
 [hep-lat/0412023].

\bibitem{tbc}
C.~T.~Sachrajda and G.~Villadoro,
  Phys.\ Lett.\ B {\bf 609}, 73 (2005)
  [hep-lat/0411033].
                                                                                        
\bibitem{Guadagnoli:2005be} 
D.~Guadagnoli, F.~Mescia and S.~Simula,
  Phys.\ Rev.\ D {\bf 73}, 114504 (2006)
  [hep-lat/0512020].

\bibitem{Boyle:2007wg} 
P.~A.~Boyle, J.~M.~Flynn, A.~J\"uttner, C.~T.~Sachrajda and J.~M.~Zanotti,
  JHEP {\bf 0705}, 016 (2007)
  [hep-lat/0703005 [HEP-LAT]].


\bibitem{MILCasqtad}
 A.~Bazavov {\it et al.},
  Rev.\ Mod.\ Phys.\  {\bf 82 } (2010)  1349-1417.
  [arXiv:0903.3598 [hep-lat]].

\bibitem{Bernard:2006nj} 
  C.~Bernard {\it et al.},  
  Phys.\ Rev.\ D {\bf 75}, 094505 (2007)
  [hep-lat/0611031].

\bibitem{Bernard:2001av} 
  C.~W.~Bernard {\it et al.}, 
  Phys.\ Rev.\ D {\bf 64}, 054506 (2001)
  [hep-lat/0104002].

\bibitem{Prelovsek05}
 S.~Prelovsek,
  Phys.\ Rev.\ D {\bf 73}, 014506 (2006)
  [hep-lat/0510080].

\bibitem{BGS06}
  C.~Bernard, M.~Golterman and Y.~Shamir,
  Phys.\ Rev.\ D {\bf 73}, 114511 (2006)
  [hep-lat/0604017].

\bibitem{Follana:2004sz}
  E.~Follana, A.~Hart and C.T.H.~Davies  [HPQCD and UKQCD Collaborations],
  Phys.\ Rev.\ Lett.\  {\bf 93} (2004) 241601
  [hep-lat/0406010].

\bibitem{Durr:2003xs} 
  S.~D\"urr and C.~Hoelbling,
  Phys.\ Rev.\ D {\bf 69}, 034503 (2004)
  [hep-lat/0311002].

\bibitem{DHW04}
S.~D\"urr, C.~Hoelbling and U.~Wagner,
Phys.\ Rev.\  D {\bf70}, 094501 (2004);

\bibitem{Donald:2011if} 
  G.~C.~Donald, C.~T.~H.~Davies, E.~Follana and A.~S.~Kronfeld,
  Phys.\ Rev.\ D {\bf 84}, 054504 (2011)
  [arXiv:1106.2412 [hep-lat]].

\bibitem{Bernard06}
 C.~Bernard,
  Phys.\ Rev.\ D {\bf 73}, 114503 (2006)
  [hep-lat/0603011].

\bibitem{Bernard:2006vv} 
  C.~Bernard, M.~Golterman, Y.~Shamir and S.~R.~Sharpe,
  Phys.\ Lett.\ B {\bf 649}, 235 (2007)
  [hep-lat/0603027].

\bibitem{Shamir04}
  Y.~Shamir,
  Phys.\ Rev.\ D {\bf 71}, 034509 (2005)
  [hep-lat/0412014]; 
  Phys.\ Rev.\ D {\bf 75}, 054503 (2007)
  [hep-lat/0607007].


\bibitem{BGS08}
 C.~Bernard, M.~Golterman and Y.~Shamir,
  Phys.\ Rev.\ D {\bf 77}, 074505 (2008)
  [arXiv:0712.2560 [hep-lat]].

\bibitem{Adams:2008db} 
  D.~H.~Adams,
  Phys.\ Rev.\ D {\bf 77}, 105024 (2008)
  [arXiv:0802.3029 [hep-lat]].


\bibitem{Allton96}
 C.~R.~Allton,
  hep-lat/9610016.

\bibitem{r1HPQCD}
  C.~T.~H.~Davies, E.~Follana, I.~D.~Kendall, G.~P.~Lepage and C.~McNeile
                  [HPQCD Collaboration],
  Phys.\ Rev.\  D {\bf 81} (2010) 034506
  [arXiv:0910.1229 [hep-lat]].

\bibitem{Btopi}
 J.~A.~Bailey {\it et al.} [Fermilab Lattice and MILC Collaboration],
 Phys.\ Rev.\  {\bf D79 } (2009)  054507.
  [arXiv:0811.3640 [hep-lat]].

\bibitem{AGtheorem}
 M.~Ademollo and R.~Gatto,
  Phys.\ Rev.\ Lett.\  {\bf 13}, 264 (1964).

\bibitem{BT03}
 J.~Bijnens and P.~Talavera,
  Nucl.\ Phys.\ B {\bf 669}, 341 (2003)
  [hep-ph/0303103].

\bibitem{Becirevic:2005py}
  D.~Becirevic, G.~Martinelli and G.~Villadoro,
  Phys.\ Lett.\ B {\bf 633}, 84 (2006)
  [hep-lat/0508013]. 

\bibitem{Bijnenspc}
Johan Bijnens, private communication. We agree with J.~Bijnens that there is a misprint in 
the $N_f=3$ formula for $f_2$ in the Appendix of Ref.~\cite{Becirevic:2005py}.


\bibitem{KtopilnuSChPT}
C.~Bernard and E.~G\'amiz, in preparation.

\bibitem{ChPTp6}
 J.~Bijnens, G.~Colangelo and G.~Ecker,
  JHEP {\bf 9902}, 020 (1999)
  [hep-ph/9902437].

\bibitem{hisqconfig}
 A.~Bazavov {\it et al.}  [MILC Collaboration],
  arXiv:1212.4768 [hep-lat].


\bibitem{Bazavov:2011fh} 
  A.~Bazavov {\it et al.}  [MILC Collaboration],
PoS LATTICE {\bf 2011}, 107 (2011) [arXiv:1111.4314 [hep-lat]].

\bibitem{LisHans}
G.~Amor\'os, J.~Bijnens and P.~Talavera,
  Nucl.\ Phys.\ B {\bf 602} (2001) 87
  [hep-ph/0101127].

\bibitem{Amoros:1999qq} 
  G.~Amor\'os, J.~Bijnens and P.~Talavera,
  Phys.\ Lett.\ B {\bf 480}, 71 (2000)
  [hep-ph/9912398].

\bibitem{L4largeNc}
J.~Gasser and H.~Leutwyler,
  Nucl.\ Phys.\ B {\bf 250}, 465 (1985). 


\bibitem{LisMILC}
A.~Bazavov {\it et al.}  [MILC Collaboration],
  PoS LATTICE {\bf 2010}, 074 (2010)
  [arXiv:1012.0868 [hep-lat]].


\bibitem{r11}
 C.~W.~Bernard {\it et al.},
  Phys.\ Rev.\ D\ {\bf 62}, 034503  (2000)
  [hep-lat/0002028].
                                
\bibitem{r12}
 R.~Sommer,
  Nucl.\ Phys.\ B\ {\bf 411}, 839  (1994)
  [hep-lat/9310022].

\bibitem{fpiMILC09}
A.~Bazavov {\it et al.} [MILC Collaboration],
  PoS\ {\bf CD09}, 007  (2009)
  [arXiv:0910.2966 [hep-ph]].

\bibitem{PDG2011}
K.~Nakamura {\it et al.} [Particle Data Group],
  J.\ Phys.\ G {\bf G37}, 075021 (2010).

\bibitem{decayconstants2011}
  A.~Bazavov {\it et al.} [Fermilab Lattice and MILC Collaborations],
  arXiv:1112.3051 [hep-lat].


\bibitem{Ciriglianof+}
V.~Cirigliano, G.~Ecker, M.~Eidem\"uller, R.~Kaiser, A.~Pich and J.~Portol\'es,
  JHEP {\bf 0504}, 006 (2005)
  [hep-ph/0503108].


\bibitem{Jaminf+}
 M.~Jamin, J.~A.~Oller and A.~Pich,
  JHEP {\bf 0402}, 047 (2004)
  [hep-ph/0401080].

\bibitem{LR84}
 H.~Leutwyler and M.~Roos,
  Z.\ Phys.\ C {\bf 25}, 91 (1984).
              
\bibitem{Kastner:2008ch} 
  A.~Kastner and H.~Neufeld,
  Eur.\ Phys.\ J.\ C {\bf 57}, 541 (2008)
  [arXiv:0805.2222 [hep-ph]].


\bibitem{tauVus}
E.~G\'amiz, M.~Jamin, A.~Pich, J.~Prades and F.~Schwab,
  Phys.\ Rev.\ Lett.\  {\bf 94}, 011803 (2005)
  [hep-ph/0408044]; 

 E.~G\'amiz, M.~Jamin, A.~Pich, J.~Prades and F.~Schwab,
  Conf.\ Proc.\ C {\bf 060726}, 786 (2006)
  [hep-ph/0610246].

\bibitem{Passemartau2012}
V.~Bernard, D.~R.~Boito and E.~Passemar,
  Nucl.\ Phys.\ Proc.\ Suppl.\  {\bf 218}, 140 (2011)
  [arXiv:1103.4855 [hep-ph]].


\bibitem{fK1}
C.~T.~H.~Davies, C.~McNeile, E.~Follana, G.~P.~Lepage, H.~Na and J.~Shigemitsu [HPQCD Collaboration],
  Phys.\ Rev.\ D {\bf 82}, 114504 (2010)
  [arXiv:1008.4018 [hep-lat]].

\bibitem{fK2}
 C.~Kelly [RBC and UKQCD Collaborations],
  PoS LATTICE {\bf 2011}, 285 (2011)
  [arXiv:1201.0706 [hep-lat]].


\bibitem{Bae:2010ki}
  T.~Bae, Y.~-C.~Jang, C.~Jung, H.~-J.~Kim, J.~Kim, K.~Kim, W.~Lee and S.~R.~Sharpe,  
  Phys.\ Rev.\ D {\bf 82}, 114509 (2010)
  [arXiv:1008.5179 [hep-lat]].


\bibitem{Bijnens:2002hp}
  J.~Bijnens and P.~Talavera,
  JHEP {\bf 0203}, 046 (2002)
  [hep-ph/0203049].

\bibitem{SCHPT}
C.~Aubin and C.~Bernard,
  Phys.\ Rev.\ D {\bf 68}, 034014 (2003)
   [arXiv:hep-lat/0304014]; Phys.\ Rev.\ D {\bf 68}, 074011 (2003)
   [arXiv:hep-lat/0306026].


\end{thebibliography}
\end{document}